\documentclass[12pt]{iopart}

\usepackage{amssymb}
\usepackage{tensor}
\usepackage{mathrsfs}
\usepackage{braket}
\usepackage{xfrac}
\usepackage{color}
\usepackage{graphicx}
\usepackage{cite}

\newcommand{\vc}[1]{\ensuremath{\mathbf{#1}}}

\newcommand{\ham}{\ensuremath{\mathcal{H}}}
\newcommand{\csm}{\ensuremath{\mathcal{M}}}
\newcommand{\ct}{\ensuremath{\tau}}

\newcommand{\adm}{\textsc{adm}}
\newcommand{\flrw}{\textsc{flrw}}
\newcommand{\cmb}{\textsc{cmb}}
\newcommand{\wkb}{\textsc{wkb}}
\newcommand{\A}{S}
\newcommand{\B}{I}
\newcommand{\Epsilon}{\mathcal{E}}

\newcommand{\sd}{\ensuremath{\mathrm{D}}}
\newcommand{\bd}{\sd_t}
\newcommand{\chubble}{\ensuremath{H}}

\begin{document}
\title[Quantum gravitational corrections to the inflationary power spectra \ldots]
 {Quantum gravitational corrections to the inflationary power spectra in scalar-tensor theories}
\author{Christian F Steinwachs and Matthijs L van der Wild}
\address{Physikalisches Institut, 
 		Albert-Ludwigs-Universit\"at Freiburg,
 		Hermann-Herder-Str. 3, 
 		79104,
 		Freiburg im Breisgau,
 		Germany}
\eads{\mailto{christian.steinwachs@physik.uni-freiburg.de},\\\hspace{1.1cm} \mailto{wild@physik.uni-freiburg.de}}

\begin{abstract}
\noindent
We derive the first quantum gravitational corrections to the inflationary power spectra for a general single-field scalar-tensor theory which includes a non-minimal coupling to gravity, a non-standard scalar kinetic term and an arbitrary potential of the scalar field. We obtain these corrections from a semiclassical expansion of the Wheeler-DeWitt equation, which, in turn, governs the full quantum dynamics in the canonical approach to quantum gravity.  
We discuss the magnitude and relevance of these corrections, as well as their characteristic signature in the inflationary spectral observables. 
We compare our results to similar calculations performed for a minimally coupled scalar field with a canonical kinetic term and discuss the impact of the non-minimal coupling on the quantum gravitational corrections.
\end{abstract}

\maketitle

 		
\section{Introduction}\label{Sec:Intro}
The inflationary paradigm is an integral part in the dynamics of the early universe and explains the formation of structure out of tiny quantum fluctuations \cite{Starobinsky1980,Starobinsky1982,Guth1981,Linde1982,Albrecht1982,Starobinsky1979,Mukhanov1981,Mukhanov1982,Guth1982,Hawking1982,Bardeen1983}.
There is a plethora of inflationary models leading to different theoretical predictions that can be tested with observational data \cite{Martin2014}.
The predictions for the inflationary power spectrum of the perturbations can be tested by analyzing the temperature anisotropy spectrum of the cosmic microwave background radiation (\cmb{}) as measured by satellites such as \textit{Planck} \cite{Akrami2018}.   

In the theoretical description of the inflationary mechanism, the quasi-De Sitter phase of accelerated expansion is realized usually by one or more scalar fields.
In the simplest models a single scalar field, the inflaton, which is minimally coupled to gravity with a canonical kinetic term, drives inflation \cite{Linde1983}.
In this case, the inflationary predictions within the slow-roll approximation are entirely determined by the scalar field potential.
However, recent observational data favors inflationary models based on more general scalar-tensor theories \cite{Brans1961,Dicke1962,Spokoiny1984,Salopek1989,Fakir1990,Sotiriou2006,Nojiri2017}, as well as geometric modifications of gravity such as $f(R)$ theories \cite{Buchdahl1970,Starobinsky1980, DeFelice2010}.
Two prominent representatives of these two classes are the model of Higgs inflation, in which the inflaton is non-minimally coupled to gravity and identified with the Standard Model Higgs boson \cite{Bezrukov2008}, and Starobinsky's $R+R^2$ model of inflation in which the inflaton is identified with the scalaron \cite{Starobinsky1980}, the additional propagating geometrical scalar degree of freedom that arises effectively due to the presence of higher derivatives in the quadratic curvature invariant \cite{Stelle1978}.
Both models lead to almost indistinguishable predictions for the inflationary spectral observables \cite{Barvinsky2008,Bezrukov2012,Barvinsky2012,Kehagias2014}.
They are both representatives of a larger class of inflationary attractors \cite{Mukhanov2013,Kallosh2014}.
Moreover, these similarities are a manifestation of a more general equivalence between scalar-tensor theories and $f(R)$ gravity, see e.g. \cite{Sotiriou2006}. Based on the one-loop results \cite{Ruf2018} obtained within the perturbative covariant approach, this equivalence was shown recently to also hold at the quantum level \cite{Ruf2018a}, see also \cite{Ohta2018} for a similar analysis.
These questions are closely related to the question of quantum equivalence between different field parametrizations in scalar-tensor theories \cite{Kamenshchik2015} for which the one-loop corrections have been obtained in \cite{Barvinsky1993,Shapiro1995,Steinwachs2011}. 

Recently, it was suggested to combine these two models into a single scalar-tensor theory \cite{Salvio2015,Kaneda2016, Ema2017a,Wang2017,Mori2017,Pi2018,Enckell2018,He2018,Gorbunov2018,Ghilencea2018,Gundhi2018,Karam2019,Canko2019,Vicentini2019, Bezrukov2019}, dubbed ``scalaron-Higgs inflation'' in \cite{Gundhi2018};  see also \cite{Enckell2019,Tenkanen2019,Antoniadis2019} for a similar analysis within the Palatini formalism.
At the classical level, this combined model features an asymptotic scale invariance, which is the theoretical motivation for many interesting models, see e.g. \cite{Meissner2007,Shaposhnikov2009,Blas2011b,Salvio2014,Shaposhnikov2018,Shaposhnikov2018a,Wetterich2019}.

Quantum corrections in these inflationary models can become important. 
For example, in the model of Higgs-inflation, the radiative corrections and the renormalization group improvement turned out to be crucial for the consistency with particle physics experiments \cite{Barvinsky2008,Bezrukov2009a,DeSimone2009,Barvinsky2009,Bezrukov2009,Bezrukov2011, Barvinsky2012,Allison2014}.
While in this case the quantum corrections are dominated by the heavy Standard Model particles, it is in general interesting to also study the effect of quantum gravitational corrections on inflationary predictions.
In fact, the strong curvature regime during the inflationary phase make the early universe a natural testing ground for any theory of quantum gravity.

In this paper, we calculate the first quantum gravitational corrections to the inflationary power spectra obtained by a canonical quantization of a general scalar-tensor theory in the framework of quantum geometrodynamics, which is one of the earliest attempts of a non-perturbative quantization of gravity \cite{Wheeler1957}. See \cite{Kiefer2012} for a comprehensive overview about various approaches to quantum gravity.

In the canonical approach spacetime is foliated into leaves of spatial hypersurfaces and the spatial metric as well as its conjugated momentum are the natural variables to be quantized.
Making use of the Arnowitt-Deser-Misner  (\adm{}) formalism, the covariant action is cast into a constrained Hamiltonian system \cite{Arnowitt1960}.
The Dirac quantization of the Hamiltonian constraint, which governs the dynamics of this system, leads to the Wheeler-DeWitt equation \cite{Dirac1964,DeWitt1967}. 
Although the canonical approach to quantum gravity does not come without difficulties, both at the conceptual and technical level, the Wheeler-DeWitt equation can be considered as a natural starting point for the analysis of quantum gravitational effects, as its semiclassical expansion reproduces the classical theory and the functional Schr\"odinger equation for quantized matter fields on a curved background at the lowest orders of the expansion \cite{Kiefer1991,Kiefer1994}.
Therefore, higher order terms in the expansion can be clearly attributed to the first quantum gravitational corrections.
When applied to the inflationary universe, these quantum gravitational corrections leave observational signatures in the primordial power spectrum. This has been investigated for a minimally coupled scalar field with a canonical kinetic term \cite{Kiefer2012a, Bini2013, Kamenshchik2013, Kamenshchik2014a, Brizuela2016, Brizuela2016a, Brizuela2019, Brizuela2019a}.
In this work we generalize these analyses to a general scalar-tensor theory of a single scalar field with an arbitrary non-minimal coupling to gravity, a non-standard kinetic term, and an arbitrary scalar potential.  

The paper is structured as follows:
in Sec. \ref{sec:inflation}, we introduce the model, derive the equations of motion, perform the reduction to a homogeneous and isotropic Friedmann-Lema\^\i tre-Robertson-Walker (\flrw{}) universe and discuss the inflationary dynamics of the background and the cosmological perturbations.
In Sec. \ref{sec-QGD}, we derive the classical Hamiltonian constraint, perform the Dirac quantization, and obtain the Wheeler-DeWitt equation for the background and perturbation variables.
In Sec \ref{sec:semiclassical-expansion}, we perform a semiclassical expansion based on a combined Born-Oppenheimer and Wentzel-Kramers-Brillouin (\wkb)-type approximation.
At the lowest orders we recover the dynamical background equations and the notion of a semiclassical time. At the next order, we obtain a Schr\"odinger equation for the perturbations. Finally, the subsequent order yields the first quantum gravitational corrections to the Schr\"odinger equation.
In Sec. \ref{sec:qg-corrections}, we derive the connection between the results found from the semiclassical expansion of the Wheeler-DeWitt equation and the inflationary power spectra.
In Sec. \ref{sec:observational-signatures}, we discuss the impact of the leading quantum gravitational corrections on the inflationary power spectra and their observational consequences.
Finally, we summarize our main results and conclude in Sec. \ref{sec:conclusion}.
The results for the subleading slow-roll contributions to the quantum gravitational corrections are provided in \ref{Appendix}.

\section{Scalar-tensor theories of inflation}
\label{sec:inflation}

Almost all models of inflation driven by a single scalar field $\varphi$ can be covered by the general scalar-tensor theory
\begin{equation}
\label{sec-infldyn:action}
S[g,\varphi] = \int\mathrm{d}^4X\sqrt{-g}\left[UR - \frac{1}{2} G\,g^{\mu\nu}\nabla_\mu\varphi\nabla_\nu\varphi - V\right].
\end{equation}
Here, $U(\varphi)$, $G(\varphi)$, and $V(\varphi)$ are three arbitrary functions of the scalar inflaton field $\varphi$. 
They parametrize the non-minimal coupling to gravity, 
the non-canonical kinetic term, and the scalar field potential, respectively. 
We work in four dimensional spacetime with metric $g_{\mu\nu}(X)$ of mostly plus signature.
The scalar curvature is denoted by $R(g)$.
Spacetime coordinates are labeled by $X^{\mu}$, $\mu=0,\ldots,3$, with capital letters.

\subsection{Field equations and energy-momentum tensor}
The field equations for the metric and the Klein-Gordon equation of the inflaton are obtained by varying \eref{sec-infldyn:action} with respect to $g_{\mu\nu}$ and  $\varphi$, respectively:
\begin{eqnarray}
\label{sec-infldyn:scalar-eom}
R_{\mu\nu}-\frac{1}{2} R g_{\mu\nu} = \frac{1}{2U}T_{\mu\nu}^{\varphi},\\
\Box\varphi = \frac{1}{G}\left(V_1-\frac{1}{2}G_1\nabla_\mu\varphi\nabla^\mu\varphi-U_1R\right),
\label{sec-infldyn:einstein-eom}
\end{eqnarray}
where the effective energy-momentum tensor $T_{\mu\nu}^{\varphi}$ is defined as
\begin{equation}
\label{eq:energy-momentum}
T_{\mu\nu}^\varphi:= G \left(\delta_\mu^\alpha\delta_\nu^\beta-\frac{1}{2} g_{\mu\nu}g^{\alpha\beta}\right)\nabla_\alpha\varphi\nabla_\beta\varphi-g_{\mu\nu}V+2\nabla_\mu\nabla_\nu U-2g_{\mu\nu}\Box U.
\end{equation}
Here $\Box:= g^{\mu\nu}\nabla_{\mu}\nabla_{\nu}$ denotes the covariant d'Alembert operator.
    
\subsection{Spacetime foliation}
It is useful to reformulate the action in terms of the \adm{} formalism \cite{Arnowitt1960}, where the four dimensional metric $g_{\mu\nu}$ is expressed in terms of the lapse function $N(t,\mathbf{x})$, the spatial shift vector $N^a(t,\mathbf{x})$, and the spatial metric $\gamma_{ab}(t,\mathbf{x})$,
\begin{equation}
\label{ADMLineElement}
\mathrm{d}s^2 = g_{\mu\nu}\mathrm {d}X^{\mu}\mathrm {d}X^{\nu}
              = -N^2\mathrm{d}t^2+2N_{a}\mathrm{d}t\mathrm{d}x^{a}+\gamma_{ab}\mathrm{d}x^{a}\mathrm{d}x^{b}.
\end{equation}
Here, the spatial coordinates $x^{a}$, $a=1,\ldots,3$ are denoted by small letters. 
In terms of the \adm{} variables, the action \eref{sec-infldyn:action} can be compactly represented as
\begin{equation}
\label{eq:background-action}
S[\gamma,\varphi] = \int \mathrm{d} t\mathrm{d}^3x\left[\frac{1}{2}\csm_{AB} \mathrm{D}_{t} \mathcal{Q}^A \mathrm{D}_{t} \mathcal{Q}^B - \mathcal{P}\right],
\quad  
\mathcal{Q}^{A} = \left(\begin{array}{c}\gamma_{ab}\\\varphi\end{array}\right),
\end{equation}
with the dynamical configuration space variables collectively denoted by $\mathcal{Q}^{A}$
and the reparametrization invariant covariant time derivative $\mathrm{D}_{t}:=(\partial_t-\mathcal{L}_{\vc{N}})/N$,
with the Lie derivative $\mathcal{L}_{\vc{N}}$ along the spatial shift vector $\vc{N}=N^a\partial_a$. 
The bilinear form $\mathcal{M}^{AB}$ corresponds to the inverse of the configuration space metric derived in \cite{Steinwachs2018},
\begin{equation}
\label{CSM}
\mathcal{M}_{AB} = -N\gamma^{1/2}
                    \left(
                    	\begin{array}{cc}
                     	-\frac{U}{2}G^{abcd}&U_{1}\gamma^{ab}\\
                    	 U_{1}\gamma^{cd}&-G
                    	 \end{array}
                    \right),
\end{equation}
with the DeWitt metric $G^{abcd}:=\gamma^{a(c}\gamma^{d)b}-\gamma^{ab}\gamma^{cd}$.
\footnote{Note that the configuration space metric \eref{CSM} was defined with additional inverse factors of the lapse function in \cite{Steinwachs2018}.}
The potential $\mathcal{P}$, which includes the spatial gradient terms of the scalar field, is defined as
\begin{equation}
\mathcal{P}:= N\gamma^{1/2}\left[\frac{1}{2} s^{-1}\sd_a\varphi \sd^a\varphi+V-UR^{(3)}-2\Delta U-\frac{3}{2}U^{-1}\sd_aU\sd^aU\right].
\end{equation}
Here, $\Delta := -\gamma^{ab}\sd_a\sd_b$ is the positive definite spatial Laplacian, $\sd_a$ the spatial covariant derivative compatible with $\gamma_{ab}$ and $R^{(3)}$ is the three-dimensional spatial curvature. 
In addition, we have defined the suppression function \cite{DeSimone2009,Barvinsky2009,Steinwachs2011,Steinwachs2018}:
\begin{equation}
\label{sec-infldyn:sfunction}
s := \frac{U}{GU+3U_1^2}.
\end{equation}
The subscript is a shorthand for a derivative with respect to the argument, 
i.e. we denote the derivative of a general field-dependent scalar function $f(\varphi)$ with respect to $\varphi$ as
\begin{equation}
 f_n(\varphi) := \frac{\partial^nf(\varphi)}{\partial \varphi^n}.
\end{equation}

\subsection{Cosmological background evolution}
\label{sec-CosmologicalBackgroundEvolution}
The background spacetime is described by a spatially flat ($R^{(3)}=0$) \flrw{} line element 
\begin{equation}
\mathrm{d} s^2 = -N^2\,\mathrm{d} t^2+a^2\delta_{ab}\,\mathrm{d} x^a\mathrm{d} x^b. \label{FRWLineelement}
\end{equation}
Comparing with the \adm{} line element \eref{ADMLineElement}, spatial flatness, homogeneity and isotropy imply $\gamma_{ab}=a^2\delta_{ab}$ and $N_{a}=0$, where the lapse function $N=N(t)$ and the scale factor \(a=a(t)\) are functions of time $t$ only. 
Similarly, homogeneity implies that the scalar field is a function of time only $\varphi=\varphi(t)$.
Moreover, for the isotropic line element \eref{FRWLineelement} the reparametrization invariant time derivative $\bd$ reduces to $\bd=N^{-1}\partial_t$. 
In addition, it is convenient to introduce the conformal time $\ct$, related to the coordinate time $t$ by $N\mathrm{d}t=a\mathrm{d}\ct$. 
This choice corresponds to the gauge $N=a$, which we will adopt in what follows. 
In terms of $\ct$, the \flrw{} metric \eref{FRWLineelement} acquires the manifestly  conformally flat structure $g_{\mu\nu}(\ct)=a^2(\ct)\eta_{\mu\nu}$, and the reparametrization invariant covariant time derivative reduces to a partial derivative with respect to conformal time $\bd=a^{-1}\partial_{\ct}$.
The conformal Hubble parameter $\chubble(\ct)$ is defined as 
\begin{equation}
\chubble := \frac{a^{\prime}}{a},
\end{equation}
where the prime denotes a derivative with respect to conformal time $\ct$.
In the \flrw{} universe $T_{\mu\nu}^\varphi$ takes on the form of the energy-momentum tensor of a perfect fluid:
\begin{equation}
T_{\mu\nu}^\varphi = (\rho_\varphi + p_\varphi) u_\mu u_\nu + p_\varphi\, g_{\mu\nu}.
\end{equation}
Here, $u_\mu$ is the fluid's four-velocity with norm $u_\mu u^\mu=-1$, $\rho_\varphi$ is its energy density, and $p_\varphi$ is its pressure.
Comparison with \eref{eq:energy-momentum} leads to the identifications
\begin{eqnarray}
\rho_\varphi =& \frac{G}{2a^2}\, (\varphi^{\prime})^2+V-\frac{6U^{\prime}}{a^2}\chubble,\label{eq:energy}\\
p_\varphi =& \frac{G}{2a^2}\, (\varphi^{\prime})^2 - V +\frac{2 U^{\prime\prime}}{a^2}+\frac{2U^{\prime}}{a^2}\chubble.\label{eq:pressure}
\end{eqnarray}
The symmetry reduced \flrw{} action expressed in terms of the compact notation
has a form similar to \eref{eq:background-action} with $\mathcal{Q}^A=(a,\varphi)$ and 
\begin{equation}
\label{eq:minisuperspace-metric-potential}
\csm_{AB}=-
\left(
	\begin{array}{cc}
	12U    & 6U_1a \\
	6U_1a & -Ga^2
	\end{array}
\right),\qquad \mathcal{P} = a^4V.
\end{equation}
The explicit expression for the background action is given by
\begin{eqnarray}
\label{BGact}
S^{\mathrm{bg}}[a,\varphi] = \int\mathrm{d}\ct\mathrm{d}^3x\mathcal{L}^{\mathrm{bg}}\left(a,a^{\prime},\varphi,\varphi^{\prime}\right),\\
\mathcal{L}^{\mathrm{bg}}(a,a^{\prime},\varphi,\varphi^{\prime}) := a^4\left[-\frac{6U}{a^2}\left(\frac{a^{\prime}}{a}\right)^2-\frac{6U_1}{a}\frac{\varphi^{\prime}a^{\prime}}{a^2}+\frac{G}{2}\left(\frac{\varphi^{\prime}}{a}\right)^2-V\right].
\label{BGlag}
\end{eqnarray}
In particular, the derivative coupling between the gravitational and scalar field degrees of freedom induced by the non-minimal coupling becomes manifest. 
The Friedmann equations and the Klein-Gordon equation are obtained from varying \eref{BGact} with respect to $N$, $a$, and $\varphi$, or directly from symmetry reducing the equations of motion \eref{sec-infldyn:scalar-eom} and \eref{sec-infldyn:einstein-eom}:
\begin{eqnarray}    
\label{eq:friedmann-equation}
&\chubble^2= \frac{1}{6}a^{2}\, U^{-1}\rho_\varphi  ,\\
\label{eq:eom-hubble}
&\chubble^{\prime}= - \frac{1}{12}a^{2}\,U^{-1}\left(\rho_\varphi + 3p_\varphi\right),\\
\label{eq:reduced-scalar-eom}
& \varphi'' + 2\chubble\varphi' + \frac{1}{2} \frac{(U/s)'}{U/s}\varphi' + a^2sU^2W_1 = 0.
\end{eqnarray}
The dimensionless ratio $W$, related to the Einstein frame potential  \cite{Steinwachs2018}, is defined as
\begin{equation}
W:=\frac{V}{U^2}.
\end{equation}
In a flat spatially homogeneous \flrw{} universe, the spatial integral in the action \eref{BGact} is formally divergent, corresponding to an infinite spatial volume $V_0$. 
In order to regularize the spatial integral, we need to introduce some large but finite reference length scale $\ell_0$ such that $\ell^3_0=V_0$.
The reference volume $V_0$ can be removed from the formalism by absorbing it into a redefinition of the time variable and the scale factor, such that the action \eref{BGact} is independent of $V_0$ \cite{Kamenshchik2014a,Brizuela2016}:
\begin{eqnarray}
\label{rescaling-background}
\ct&\to\ell_0^{-1}\ct,\qquad a&\to\ell_0a.\label{rescal1}
\end{eqnarray}
While in this way any dependence on the reference scale $\ell_0$ has been eliminated form the formalism, the restriction of the spatial volume to a compact subregion nevertheless has observational consequences, which we discuss in Sec. \ref{sec:observational-signatures}.

\subsection{Inflationary background dynamics in the slow-roll approximation }
\label{sec:slow-roll}
During inflation the universe undergoes a quasi-De Sitter stage in which the energy density is approximately constant and effectively dominated by the potential of the slowly rolling scalar field. 
In scalar-tensor theories, the slow-roll conditions can be generalized for any function $f$ of the scalar field $\varphi$ as \cite{Torres1997}: 
\begin{equation}
\label{eq:slow-roll-def}
f^{\prime\prime}(\varphi)\ll\chubble f^{\prime}(\varphi)\ll\chubble^2f(\varphi).
\end{equation}
In particular, for the scalar-tensor theory \eref{sec-infldyn:action}, this encompasses the generalized potentials $f=\{U,\,G,\,V\}$.
Making use of \eref{eq:energy} and \eref{eq:pressure}, within the slow-roll regime \eref{eq:slow-roll-def} the background equations \eref{eq:friedmann-equation}-\eref{eq:reduced-scalar-eom} lead to
\begin{eqnarray}
\frac{\chubble^2}{Ua^2}\approx{\frac{W}{6}},\qquad\label{eq:attractor2}
3\frac{\chubble\varphi^{\prime}}{U^2a^2}\approx-sW_1\,.\label{eq:attractor1}
\end{eqnarray}
The slow-roll conditions \eref{eq:slow-roll-def} motivate the definitions of the following four slow-roll parameters \cite{Hwang1997}, which quantify small deviations from De Sitter space:
\begin{eqnarray}
\varepsilon_{1,\chubble} := 1 - \frac{\chubble'}{\chubble^2},\qquad
&\varepsilon_{2,\chubble}:= 1 - \frac{\varphi''}{\chubble\varphi'},\nonumber\\
\varepsilon_{3,\chubble} := \frac{1}{2}\frac{U'}{\chubble U},\qquad
&\varepsilon_{4,\chubble}:= \frac{1}{2}\frac{(U/s)'}{\chubble(U/s)}.
\label{HubbleSLP}                   
\end{eqnarray}
The slow-roll parameters $\varepsilon_{1,\chubble}$ and $\varepsilon_{2,\chubble}$ are the same as for a minimally coupled canonical scalar field, 
while the slow-roll parameters $\varepsilon_{3,\chubble}$, $\varepsilon_{4,\chubble}$ contain
information about the non-minimal coupling $U$ and the generalized kinetic term $G$ via the function $s$ defined in \eref{sec-infldyn:sfunction}.
We work to linear order in the slow-roll approximation, where the $\varepsilon_{i,\chubble}$, $i=1,\ldots,4$ are treated as constant. 
In addition to \eref{HubbleSLP}, we define another set of ``potential'' slow-roll  parameters $\varepsilon_{i,W}$, which are expressed directly in terms of the generalized potentials $U$, $G$ and $V$ and their derivatives,
\begin{eqnarray}
\varepsilon_{1,W} := \frac{(UW)_1}{UW}\frac{sUW_1}{W},\qquad
&\varepsilon_{2,W} := 2\left(\frac{sUW_1}{W}\right)_1+\frac{sUW_1}{W}\frac{(UW)_1}{UW},\nonumber\\
\varepsilon_{3,W} := -\frac{sU_1W_1}{W},\qquad
&\varepsilon_{4,W} := -\frac{sUW_1}{W}\frac{(U/s)_1}{U/s}.
\label{PotSLP}
\end{eqnarray}
Within the slow-roll approximation $ \varepsilon_{i,W}\approx \varepsilon_{i,\chubble}$. Therefore, in what follows we simply write $\varepsilon_{i}$ for both sets of slow-roll parameters \eref{HubbleSLP} and \eref{PotSLP}.
During the slow-roll regime, a sufficiently long quasi-De Sitter phase of inflation is realized for $|\varepsilon_i|\ll1$.

\subsection{Cosmological perturbations}
\label{sec:quantum-cosmology}
We split the fields $g_{\mu\nu}$ and $\varphi$ into background $\bar{g}_{\mu\nu}$, $\bar{\varphi}$ and perturbation $\delta g_{\mu\nu}$ and $\delta\varphi$, 
\begin{equation}
\label{perts}
g_{\mu\nu}(\,\mathbf{x}):=\bar{g}_{\mu\nu}(\ct)+\delta g_{\mu\nu}(\ct,\mathbf{x}),\qquad 
\varphi(\ct,\mathbf{x}):= \bar{\varphi}(\ct)+\delta\varphi(\ct,\mathbf{x}).
\end{equation}
In the following, we omit the bar. 
In addition, we decompose the metric perturbation $\delta g_{\mu\nu}$ into its irreducible components. 
The cosmological line element including linear perturbations can be parametrized as
\begin{equation}
\label{eq:perturbed-line-element}
\mathrm{d} s^2 = a^2(\ct)\left[-(1+2n)\mathrm{d}\ct^2 + 2n_a\mathrm{d} \ct\mathrm{d} x^a + \left(\delta_{ab}+2h_{ab}\right)\mathrm{d}x^a\mathrm{d}x^b\right]. 
\end{equation}
Here, $n$ is the scalar perturbation of the lapse function. 
The perturbation of the shift vector $n_a(t,\mathbf{x})$, as well as the spatial metric $h_{ab}(t,\mathbf{x})$, are further decomposed as
\begin{equation}   
n_a  =\partial_a n^{\mathrm{L}} +n^{\mathrm{T}}_a,\qquad 
h_{ab} = \left[ \phi\,\delta_{ab} + \partial_a\partial_b h^{\mathrm{L}} + \partial_{(a}h^{\mathrm{T}}_{b)} + h^{\mathrm{TT}}_{ab}\right],
\end{equation}
with the scalar perturbations $n^{\mathrm{L}}$, $\phi$, $h^{\mathrm{L}}$, 
the transverse vector perturbations ${\partial^{a}n^{\mathrm{T}}_a=\partial^{a}h^{\mathrm{T}}_{a}=0}$, 
and the transverse traceless tensor perturbation {$\partial^{a}h^{\mathrm{TT}}_{ab}=\delta^{ab}h^{\mathrm{TT}}_{ab}=0$}. 
Since vector modes decay during inflation, 
we neglect them in what follows and focus on the scalar and tensor perturbations.
The gauge invariant transverse traceless part $h_{ab}^{\mathrm{TT}}$ of the metric perturbation can be associated with primordial gravitational waves
\begin{equation}
h^{\mathrm{TT}}_{ab}:=\sum_{I=+,\times}e_{ab}^{I}h^{\mathrm{TT}}_I,
\end{equation}
where $e_{ab}^{I}$ denotes the polarization tensor. 
Since the scalar perturbations are gauge dependent, it is convenient to work with the single scalar gauge invariant combination, the Mukhanov-Sasaki variable \cite{Mukhanov1992}:
\begin{equation}
\delta\varphi_{\mathrm{g}} := \delta\varphi - \frac{\varphi'}{\chubble} \phi.
\end{equation}
Finally, we introduce the canonical field variables for the scalar perturbation $\delta\varphi_{\mathrm{g}}$ and the transverse-traceless tensor perturbation $h_{ab}^{\mathrm{TT}}$:
\begin{equation}
\label{eq:mukhanov-sasaki}
v    := a\, z_{\mathrm{S}}\,\delta\varphi_{\mathrm{g}},\qquad
u_{I}:= a\, z_\mathrm{T}\, h^{\mathrm{TT}}_I.
\end{equation}
The corresponding factors $z_{\mathrm{S}}$ and $z_{\mathrm{T}}$ defined as \cite{Hwang1997}:
\begin{equation}
\label{eq:z}
z_{\mathrm{S}}^2:= s^{-1}\left(1+\frac{1}{2\chubble}\frac{ U'}{ U}\right)^{-2}\left(\frac{\varphi'}{\chubble}\right)^2,
\qquad
z_{\mathrm{T}}^2   := \frac{U}{2}.
\end{equation}
The action quadratic in the perturbations $v$ and $u_I$ reads \cite{Hwang1997}:
\begin{eqnarray}
S^{\mathrm{pert}}[v,u] = \int\mathrm{d}\ct\mathrm{d}^3x\left[\mathcal{L}^{S}(v,v^{\prime})+\mathcal{L}^{T}(u,u^{\prime})\right]\label{S2},\\
\mathcal{L}^{\mathrm{S}}(v,v^{\prime}) = \frac{1}{2}\left[\left(v^{\prime}\right)^2+\delta^{ij}\partial_iv\partial_jv+\frac{\left(az_{\mathrm{S}}\right)^{\prime\prime}}{\left(a z_{\mathrm{S}}\right)}v^2\right],\label{S2scalar}\\
\mathcal{L}^{\mathrm{T}}(u,u^{\prime}) =\frac{1}{2}\sum_{I=+,\times}\left[\left(u^{\prime}_{I}\right)^2+\delta^{ij}\partial_iu_{I}\partial_ju_{I}+\frac{\left(az_{\mathrm{T}}\right)^{\prime\prime}}{\left(az_{\mathrm{T}}\right)}\left(u_I\right)^2\right].\label{S2tensor}
\end{eqnarray}
In the derivation of \eref{S2scalar} and \eref{S2tensor}, total derivative terms are neglected and it is assumed that the background fields satisfy their equations of motion \eref{eq:friedmann-equation}-\eref{eq:reduced-scalar-eom}. 
Since we consider only linear perturbations, the expansion stops at second order and the total combined action of background plus perturbations reads
\begin{eqnarray}
\label{eq:def-lagrangian}
S^{\mathrm{tot}}[a,\varphi,v,u] &:=\int\mathrm{d}\ct\mathrm{d}^3x\,\mathcal{L}^{\mathrm{tot}}(a,a^{\prime},\varphi,\varphi^{\prime},v,v^{\prime},u,u^{\prime})\nonumber\\
 &\;=\int\mathrm{d}\ct\mathrm{d}^3x\left[\mathcal{L}^{\mathrm{bg}}(a,a^{\prime},\varphi,\varphi^{\prime})+\mathcal{L}^{\mathrm{S}}(v,v^{\prime})+\mathcal{L}^{\mathrm{T}}(u,u^{\prime})\right].
\label{TotAct}
\end{eqnarray}
Next, we perform a Fourier transformation of the inhomogeneous perturbations:
\begin{equation}
\label{Fourier-transform}
v(\ct,\vc{x}) = \frac{1}{V_0} \sum_{k} \rme^{\rmi\vc{k\cdot r}} v_k(\ct),
\qquad 
u(\ct,\vc{x}) = \frac{1}{V_0} \sum_{k} \rme^{\rmi\vc{k\cdot r}} u_k(\ct).
\end{equation}
Since the position space perturbations are real, we have $v_{\mathbf{k}}^{*}=v_{-\mathbf{k}}$ and $u_{\mathbf{k},I}^{*}=u_{-\mathbf{k},I}$.
The restriction of the spatial volume to a compact subregion makes it necessary to perform the discrete Fourier transform \eref{Fourier-transform} with the volume factor $V_0=\ell_0^3$, introduced to regularize the spatial integral in \eref{BGact}.
Moreover, due to the isotropy of the \flrw{} background, the mode components can only depend on the magnitude ${k:=\sqrt{\mathbf{k}\cdot\mathbf{k}}=\sqrt{\delta^{ij}k_ik_j}}$ of the wave vector $\mathbf{k}$, rather than its direction. 
The Fourier transformed action \eref{eq:def-lagrangian} then acquires the form of a sum of harmonic oscillators 
\begin{eqnarray}
S^{\mathrm{pert}}[\{v_k\},\{u_k\}] =\frac{1}{ V_0}\sum_k \int\mathrm{d}\ct\left[\mathcal{L}^{S}_k(v_k,v_k^{\prime})+\mathcal{L}^{T}_k(u_k,u_k^{\prime})\right]\label{S2k},\\
\mathcal{L}_k^{\mathrm{S}}(v_k,v_k^{\prime}) = \frac{1}{2}\left[v^{\prime}_kv^{*\prime}_k-\omega_{\mathrm{S}}^2v_kv^{*}_k\right],\label{S2kscalar}\\
\mathcal{L}^{\mathrm{T}}_k(u_k,u_k^{\prime}) =\frac{1}{2}\sum_{I=+,\times}\left[u^{\prime}_{k,I}u^{*\prime}_{k,I}-\omega^2_{\mathrm{T}}u_{k,I}u^{*}_{k,I}\right],\label{S2tensork}
\end{eqnarray}
 with time-dependent frequencies
\begin{eqnarray}
\label{perturbation-frequencies}
\omega_{\mathrm{S}}^2(\ct;k) := k^2 - \frac{(az_{\mathrm{S}})''}{az_{\mathrm{S}}},
\qquad
\omega_{\mathrm{T}}^2(\ct;k) := k^2 - \frac{(az_{\mathrm{T}})''}{az_{\mathrm{T}}}.
\end{eqnarray}
 In a similar fashion as in \eref{rescaling-background}, it is possible to eliminate any explicit occurrence of the reference volume in the Fourier transformed action \eref{S2k} by the rescalings \cite{Kamenshchik2014a,Brizuela2016},
\begin{eqnarray}
k\to\ell_0k,\qquad v_k  \to\ell^{-2}_0v_k,\qquad u_{k,I}\to\ell^{-2}_0u_{k,I}.\label{rescal2}
\end{eqnarray}
The Fourier transformed version of the total action \eref{TotAct} is the starting point for the Hamiltonian formulation carried out in the next section.

\section{Quantum Geometrodynamics}
\label{sec-QGD}
\subsection{Hamiltonian formalism}
\label{sec:hamiltonian-formalism}
The canonical quantization of gravity is based on its Hamiltonian formulation. 
We perform a Legendre transformation of $\mathcal{L}_{\mathrm{tot}}$ with the generalized momenta
\begin{eqnarray}
  \label{momenta}
  \fl
  \pi_a &:=\frac{\partial \mathcal{L}_{\mathrm{tot}}}{\partial (a^{\prime})},
  \qquad
  \pi_\varphi :=\frac{\partial \mathcal{L}_{\mathrm{tot}}}{\partial (\varphi^{\prime})},
  \qquad
  \pi_{v,k}:=\frac{\partial \mathcal{L}_{\mathrm{tot}}}{\partial (v^{*\prime}_k)},
  \qquad
  \pi_{u,k}^{I}:=\frac{\partial \mathcal{L}_{\mathrm{tot}}}{\partial (u_{k,I}^{*\prime})},
\end{eqnarray}
which leads to the Hamiltonian constraint
\begin{eqnarray}
  \label{eq:hamiltonian-constraint}
  \ham^{\mathrm{tot}} &:=\ham^{\mathrm{bg}}+\ham^{\mathrm{pert}}=0.
\end{eqnarray}
The individual Hamiltonians of the background and perturbation variables read
\begin{eqnarray}
  \ham^{\mathrm{bg}}(a,\varphi) := -\frac{s}{24Ua^2}\left(Ga^2\pi_a^2+12U_1a\pi_a\pi_\varphi-12U\pi_\varphi^2\right) + a^4 V ,\\
  \ham^{\mathrm{pert}}(v_k,u_k^{I}, a,\varphi) :=\sum_{k}\ham_{k}^{\mathrm{pert}}=\sum_k\left( \ham_{k}^{\mathrm{S}}+\ham_{k}^{\mathrm{T}}\right),\label{HPtot}\\
  \ham_{k}^{\mathrm{S}}(v_k, a,\varphi) :=\frac{1}{2}\left( \left|\pi_{v,k}\right|^2+ \omega_{\mathrm{S}}^2 \left|v_k\right|^2\right),\label{HP1}\\
  \ham_{k}^{\mathrm{T}}(u_k^{I}, a,\varphi) :=\frac{1}{2}\sum_{I=+,\times}\left( \left|\pi_{u,k}^{I}\right|^2+ \omega_{\mathrm{T}}^2\left|u_{I,k}\right|^2\right).\label{HP2}
\end{eqnarray}

\subsection{Quantum Geometrodynamics and Wheeler-DeWitt equation}    
\label{sec-QGWDW}
In the canonical quantization procedure, the configuration space variables $a$, $\varphi$, $v_k$, $u_{k,I}$, and momenta $\pi_a$, $\pi_{\varphi}$, $\pi_{v,k}$, $\pi_{u,k}^{I}$, are promoted to operators that act on states $\Psi$ and obey the canonical commutation relations (in units $\hbar\equiv1$)
\footnote{Formally, for a consistent quantization, the configuration space variables associated with the perturbations should be doubled by decomposing the complex Fourier modes $v_{k}$ and $u_{k,I}$ into real and imaginary parts \cite{Martin2012}. For the sake of a compact formulation, we proceed with quantizing terms such as $v_{k}v_{k}^{*}$ by simply treating them as $v_k^2$ -- the final results are not affected by this.},
\begin{eqnarray}
\left[\hat{a},\hat{\pi}_a\right] ={}i,&\qquad \left[\hat{\varphi},\hat{\pi}_{\varphi}\right]=i,\nonumber\\
\left[\hat{v}_k,\hat{\pi}_{v,k^{\prime}}\right] ={} i\delta_{k,k^{\prime}},&\qquad [\hat{u}_{k,I},\hat{\pi}_{u,k^{\prime}}^{J}] = i\delta_{k,k^{\prime}}\delta_{I}^{J},
\end{eqnarray}
with all other commutators equal to zero. In the Schr\"odinger representation, the position space operators act multiplicatively and the momentum space operators act as differential operators with the explicit form
\begin{equation}
\pi_a = -i\frac{\partial}{\partial a},\qquad \pi_\varphi = -i\frac{\partial}{\partial \varphi},\qquad \pi_{v,k} = -i\frac{\partial}{\partial v_k},\qquad\pi_{u,k}^{I} = -i\frac{\partial}{\partial u_{k,I}}.\label{MomOp}
\end{equation} 
The Wheeler-DeWitt equation is obtained by promoting \eref{eq:hamiltonian-constraint} to an operator equation, acting on a wave function $\Psi(a,\varphi,v_k,u_{k,I})$.
Following the prescription for the quantization of constrained systems, introduced by Dirac \cite{Dirac1964}, the implementation of the classical constraint equation \eref{eq:hamiltonian-constraint} at the quantum level corresponds to selecting only those states $\Psi$ which are annihilated by $\hat{\ham}_{\mathrm{tot}}$,
\begin{equation}
\label{eq:wheeler-dewitt}
 \hat{\ham}^{\mathrm{tot}}\Psi= 0.
\end{equation}
The Wheeler-DeWitt equation \eref{eq:wheeler-dewitt} is defined only up to operator ordering. The results for the semi-classical expansion performed in the subsequent sections are however independent of the factor ordering, see e.g. \cite{Kiefer1991, Steinwachs2018} for details.

\section{Semiclassical expansion of the Wheeler-DeWitt equation}
\label{sec:semiclassical-expansion}
For almost all cases, the full Wheeler-DeWitt equation cannot be solved exactly.
Since we are interested only in the first quantum gravitational corrections, we do not need to find exact solutions but instead perform a systematic semiclassical expansion of the Wheeler-DeWitt equation.  
This semiclassical expansion is based on the combined use of a Born-Oppenheimer and \wkb{}-type approximation scheme.
The former relies on a clear distinction between the ``heavy'' and ``light'' degrees of freedom. In the original Born-Oppenheimer approach to molecular physics, this distinction is based on the presence of a mass hierarchy between different degrees of freedom.
For a scalar field $\varphi$ minimally coupled to Einstein gravity, such a mass hierarchy could be related to the ratio $\lambda:=m_{\varphi}^2/M_{\mathrm{P}}^2\ll1$, with the effective scalar field mass $m_{\varphi}$.
In this context, the gravitational degrees of freedom are the heavy or ``slow'' ones, while the scalar field degrees of freedom are the light or ``fast'' ones \cite{Gerlach1969,Lapchinsky1979,Brout1987,Brout1989,Banks1985, Kiefer1991,Bertoni1996, Kamenshchik2013,Kamenshchik2014a}.
Such a scenario would correspond to a slowly varying background geometry on which the quantum matter (scalar field) degrees of freedom propagate.
For a scalar field non-minimally coupled to gravity, the identification of light and heavy degrees of freedom becomes more subtle \cite{Steinwachs2018}.
In the case of the Hamiltonian \eref{eq:hamiltonian-constraint}, the heavy degrees of freedom are identified with the homogeneous background variables $a$ and $\varphi$, while the light degrees of freedom are associated with the infinitely many degrees of freedom corresponding to the Fourier components of the inhomogeneous perturbations $v$ and $u^{I}$.
In the cosmological framework, this distinction follows naturally from the observed temperature anisotropies $\Delta T/T\approx10^{-5}$ in the \cmb{}.

\subsection{Implementation of the semiclassical expansion}
In the following we use a condensed notation and collectively denote the heavy degrees of freedom by $Q^{A}:=(a,\varphi)$ and the light degrees of freedom by $q_{n}:=(v_k,u^{+}_k,u^{\times}_k)$. The index $n$ labels both the Fourier modes $k$ as well as the different types of perturbations. 
At a technical level, the distinction between heavy and light degrees of freedom can be implemented by introducing a formal weighting parameter $\lambda$ in the Hamiltonian for the heavy degrees of freedom, which can be set to one after the expansion\cite{Steinwachs2018}, 
\begin{equation}
\label{eq:weighted-wheeler-dewitt-mode}
\hat{\ham}^{\mathrm{bg}}\big(\hat{Q},\hat{\pi}_Q\big)\to\hat{\ham}^{\mathrm{bg}}_\lambda\big(\hat{Q},\hat{\pi}_Q\big)=
-\frac{\lambda}{2}\csm^{AB}(\hat{Q})\frac{\partial^2}{\partial Q^A\partial Q^B} 
+ \lambda^{-1}\mathcal{P}(\hat{Q}).
\end{equation}
Here, in correspondence with the notation in \eref{eq:background-action}, $\hat{Q}^{A}$ collectively denotes the operators $\hat{a}$ and $\hat{\varphi}$, and  $\mathcal{M}^{AB}(\hat{Q})$ and $\mathcal{P}(\hat{Q})$ denote the operator versions of \eref{eq:minisuperspace-metric-potential}.
Combining \eref{eq:hamiltonian-constraint} with \eref{MomOp} and the weighting of the background Hamiltonian \eref{eq:weighted-wheeler-dewitt-mode}, the Wheeler-DeWitt equation has the form
\begin{equation}
\label{wdw}
\left(\hat{\ham}^\mathrm{bg}_\lambda+\sum_n\hat{\ham}^{\mathrm{pert}}_n\right)\Psi=0.
\end{equation}
The Hamiltonian of the perturbation $q_{n}$  has the explicit form\footnote{We are implicitly treating the background variables $a$, $\varphi$ in the frequencies $\omega_{\mathrm{S}}^2(\ct;k)$ and $\omega_{\mathrm{T}}^2(\ct;k)$ as classical, i.e. not subjected to the canonical quantization procedure. Within the semiclassical expansion, this means that the variables $Q^{A}$ enter the frequencies only parametrically via $\ct$. This procedure might be justified  a posteriori, by showing that a full quantum treatment of these variables would only affect terms at higher order in the semiclassical expansion \cite{Brizuela2016}.},
\begin{equation}
\hat{\ham}_n^{\mathrm{pert}} = \frac{1}{2}\left(-\frac{\partial^2}{\partial q_n^2} + \omega^2_n \hat{q}_n^2\right).\label{Hpert}
\end{equation}
In what follows we suppress hats on operators and resort to the abbreviated notation
\begin{equation}
\partial_A := \frac{\partial}{\partial Q^A},
\qquad
\partial_{A}\partial^{A}:= \csm^{AB}\partial_A\partial_B.
\end{equation}
The additive structure of the Wheeler-DeWitt equation \eref{wdw} suggests the product ansatz
\begin{eqnarray}
\Psi(Q,\{q_n\}) &:= \Psi_\mathrm{bg}(Q)\Psi_{\mathrm{pert}}(Q;\{q_n\}),\\
\Psi_{\mathrm{pert}}(Q;\{q_n\})&:=\prod_n\Psi_n(Q;q_n).\label{ProdAn}
\end{eqnarray}
Inserting the ansatz \eref{ProdAn} into the Wheeler-DeWitt equation \eref{wdw}, dividing the result by $\Psi$, and separating those terms which only depend on the background variables $Q$ from those that depend additionally on the perturbations $q_n$, leads to a family of  separate equations \cite{Kiefer1987, Bertoni1996}:
\begin{eqnarray}
\fl
\frac{1}{\Psi_{\mathrm{bg}}}\hat{\ham}^{\mathrm{bg}}_{\lambda}\Psi_{\mathrm{bg}}=\frac{1}{\Psi_{\mathrm{bg}}}\left[-\frac{\lambda}{2}\partial_A\partial^A + \lambda^{-1}\mathcal{P}(\hat{Q})\right]\Psi_\mathrm{bg}=f(Q),\label{wdw-background}\\
\fl \sum_n\left[-\frac{1}{2}\frac{\partial_A\partial^A\Psi_n}{\Psi_n}-\frac{\partial_A\Psi_\mathrm{bg}\partial^A\Psi_n}{\Psi_\mathrm{bg}\Psi_n}+\lambda^{-1}\frac{\ham_n\Psi_n}{\Psi_n}-\frac{1}{2}\sum_{m\neq n}\frac{\partial_A\Psi_n\partial^A\Psi_m}{\Psi_n\Psi_m}\right]=-f(Q).
\label{wdw-perturbations}
\end{eqnarray}
Here, $f(Q)$ is arbitrary function corresponding to the backreaction of the perturbations on the background. In addition, we assume the random phase approximation 
\begin{equation}
\sum_{n\neq m}\frac{\partial_A\Psi_n\partial^A\Psi_m}{\Psi_n\Psi_m}\approx0.
\end{equation}
Under these assumptions, we can write $f(Q):=\sum_nf_n(Q)$ and obtain from \eref{wdw-perturbations} a family of separate equations for each $n$. In the following we neglect these backreaction terms by choosing $f_n(Q)=0$, such that the background wave function $\Psi_{\mathrm{bg}}(Q)$ satisfies the background part of the Wheeler-DeWitt equation and \eref{wdw} reduces to the following family of equations
\begin{eqnarray}
-\frac{\lambda}{2}\partial_A\partial^A\Psi_\mathrm{bg} + \lambda^{-1}\mathcal{P}(\hat{Q})\Psi_\mathrm{bg}=0,\label{BGExp}\\
-\frac{\lambda}{2}\frac{\partial_A\partial^A\Psi_n}{\Psi_n}-\lambda\frac{\partial_A\Psi_\mathrm{bg}\partial^A\Psi_n}{\Psi_\mathrm{bg}\Psi_n}+\frac{\ham_n\Psi_n}{\Psi_n}=0.
\label{EqSemClassExp}
\end{eqnarray}
In order to proceed, we perform a \wkb-type approximation and assume that the $\Psi_{n}$ depend only adiabatically on the background variables $\Psi_n(Q,q_n)=\Psi_n(Q;q_n)$, i.e. that a change of background variables $Q$ causes the $\Psi_n$ to change much slower than $\Psi_{\mathrm{bg}}$,
\begin{equation}
\left\vert\frac{\partial_A\Psi_\mathrm{bg}}{\Psi_\mathrm{bg}}\right\vert\gg\left\vert\frac{\partial_A\Psi_n}{\Psi_n}\right\vert.\label{Adiab}
\end{equation}
This motivates the following ansatz for $\Psi_{\mathrm{bg}}$ and $\Psi_{n}$, where the expansion in $\Psi_n$ starts at order $\mathcal{O}(\lambda^0)$ rather than $\mathcal{O}(\lambda^{-1})$,
\begin{eqnarray}
\Psi_\mathrm{bg}(Q) &= \exp\left\{\rmi\left[\lambda^{-1}\A^{(0)}(Q)+\A^{(1)}(Q)+\lambda\A^{(2)}(Q)+\ldots\right]\right\},\label{ansatz-background}\\
\Psi_n(Q;q_n) &= \exp\left\{\rmi\left[\B_{n}^{(1)}(Q;q_n)+\lambda\B_{n}^{(2)}(Q;q_n)+\ldots\right]\right\}.\label{ansatz-perturbations}
\end{eqnarray}
Inserting \eref{ansatz-background} and \eref{ansatz-perturbations} into \eref{BGExp} and \eref{EqSemClassExp}, and collecting terms of equal powers in $\lambda$ leads to two families of equations for the background functions $\A^{(j)}$ and the perturbation functions $\B^{(j)}_n$ at each order in $\lambda$.
Once this set of equations has been obtained, the formal expansion parameter $\lambda$ is set to one.
The resulting equations are then solved order by order, by first solving the equations for the background as their solutions enter the equations for the perturbations.
In order to extract the first quantum gravitational corrections, it is sufficient to consider the expansions \eref{ansatz-background} and \eref{ansatz-perturbations} up to $\mathcal{O}(\lambda)$.

\subsection{Hierarchy of background equations}

By successively solving for $\A^{(0)}$, $\A^{(1)}$ and $\A^{(2)}$, we reconstruct the wave function $\Psi_\mathrm{bg}$ up to $\mathcal{O}\left(\lambda\right)$.

\subsubsection{$\mathcal{O}(\lambda^{-1})$: }
At this order, we obtain a Hamilton-Jacobi type equation for $S^{(0)}$,
\begin{equation}
\label{eq:hamilton-jacobi}
\frac{1}{2}\frac{\partial\A^{(0)}}{\partial\ct}+\mathcal{P}=0.\label{background-expansion-1}
\end{equation}
The semiclassical time $\ct$ arises from the expansion of the timeless Wheeler-DeWitt equation \eref{eq:wheeler-dewitt} as the projection along the gradient of the background geometry $\A^{(0)}(Q)$,
\begin{equation}
\label{eq:semiclassical-time}
\frac{\partial}{\partial\ct} := \partial_A\A^{(0)}\partial^A.
\end{equation}
The consistency of the semiclassical expansion requires that the classical theory is recovered at the lowest order. Indeed, by identifying the semi-classical time \eref{eq:semiclassical-time} with the conformal time $\ct$ and the gradient of $\A^{(0)}$ with the background momenta,
\begin{equation}
\label{eq:classical-momenta}
\pi_A=\frac{\partial \A^{(0)}}{\partial Q^A}, 
\end{equation}
the Hamilton-Jacobi equation \eref{eq:hamilton-jacobi} yields the equations of motion \eref{eq:friedmann-equation}-\eref{eq:reduced-scalar-eom}, upon using \eref{eq:classical-momenta}.
The Hamilton-Jacobi equation therefore implies the classical equations of motion for the background variables $Q=(a,\varphi)$.

\subsubsection{$\mathcal{O}(\lambda^{0})$: }

Equipped with the semi-classical notion of time \eref{eq:semiclassical-time}, at the next order of the semiclassical expansion we obtain
\begin{equation}
\frac{\partial\A^{(1)}}{\partial\ct}=\frac{\rmi}{2}\partial_A\partial^A\A^{(0)}\label{background-expansion-2}.\\
\end{equation}
Using the definition of the semiclassical time \eref{eq:semiclassical-time}, $\A^{(1)}$ we find
\begin{equation}
\label{van-vleck}
\A^{(1)}=-\frac{\rmi}{2}\log\Delta.
\end{equation}
Here, $\Delta$ is a function satisfying the transport equation
\begin{equation}
\partial_A\left(\Delta\partial^A\A^{(0)}\right)=0.
\end{equation}
This is consistent with the first order corrections to the \wkb{} prefactor, where $\Delta$ is associated with the Van Vleck determinant.

\subsubsection{$\mathcal{O}(\lambda^{1})$: }

At this level of the expansion we obtain
\begin{equation}
\frac{\partial\A^{(2)}}{\partial\ct}=
\frac{1}{2}\left(\rmi\,\partial_A\partial^A\A^{(1)}-\partial_A\A^{(1)}\partial^A\A^{(1)}\right).\label{background-expansion-3}
\end{equation}
Substituting \eref{van-vleck} into \eref{background-expansion-3}, we find that $\A^{(2)}$ satisfies the differential equation
\begin{equation}
\frac{\partial\A^{(2)}}{\partial\ct}=\frac{1}{4}\left[\frac{\partial_A\partial^A\Delta}{\Delta}-\frac{3}{2}\frac{\partial_A\Delta\partial^A\Delta}{\Delta^2}\right].
\end{equation}
This shows that $\A^{(2)}$ corresponds to the second order correction to the \wkb{} prefactor.

\subsection{Hierarchy of perturbation equations}

Next, we consider the expansion of \eref{EqSemClassExp}.
Using the equations \eref{background-expansion-1}-\eref{background-expansion-3} of the background equations, we reconstruct the $\Psi_n$ up to first order in the expansion parameter $\lambda$.

\subsubsection{$\mathcal{O}(\lambda^0)$: }
\label{subsubsec:schroedinger-equation}
At this order in the expansion, using \eref{eq:semiclassical-time}, we obtain
\begin{equation}
-\frac{\partial\B^{(1)}_{n}}{\partial\ct}=
-\frac{\rmi}{2}\frac{\partial^2\B^{(1)}_{n}}{\partial q^2_n}+\frac{1}{2}\frac{\partial\B^{(1)}_{n}}{\partial q_n}\frac{\partial\B^{(1)}_{n}}{\partial q_n}+\frac{1}{2}\omega_n^2q_n^2.\label{perturbations-expansion-1}
\end{equation}
This equation is equivalent to the Schr\"odinger equation for the states $\Psi^{(1)}_n:=\exp(\rmi\B^{(1)}_{n})$,
\begin{equation}
\label{schroedinger-equation}
\rmi\frac{\partial\Psi^{(1)}_n}{\partial\ct}=\hat{\ham}_n^{\mathrm{pert}}\Psi^{(1)}_n.
\end{equation}

\subsubsection{$\mathcal{O}(\lambda^{1})$: }
\label{subsubsec:gqc-schroedinger-equation}
The first quantum gravitational corrections arise from the semiclassical expansion at order $\mathcal{O}(\lambda^{1})$. Making use of \eref{eq:semiclassical-time}, we obtain
\begin{eqnarray}
\fl-\frac{\partial\B^{(2)}_{n}}{\partial\ct}=\partial_A\A^{(1)}\partial^A\B^{(1)}_{n}+\frac{1}{2}\partial_A\B^{(1)}_{n}\partial^A\B^{(1)}_{n}-\frac{\rmi}{2}\partial_A\partial^A\B^{(1)}_{n}-\frac{\rmi}{2}\frac{\partial^2\B^{(2)}_{n}}{\partial q_n^2}+\frac{\partial\B^{(1)}_{n}}{\partial q_n}\frac{\partial\B^{(2)}_{n}}{\partial q_n}.\label{perturbations-expansion-2}
\end{eqnarray}
This equation can be written in the form of a corrected Schr\"odinger equation for the state $\Psi^{(2)}_n:=\Psi^{(1)}_n\exp(\rmi\lambda\B^{(2)}_{n})$, with $\Psi^{(1)}_n$ satisfying \eref{schroedinger-equation}\footnote{In the transition from \eref{perturbations-expansion-2} to \eref{pertexppsi}, terms of order $\mathcal{O}\left(\lambda^2\right)$ are neglected when converting derivatives of $I^{(2)}$ in terms of derivatives of $\Psi^{(2)}$.}, 
\begin{equation}
\label{pertexppsi}
\rmi\frac{\partial\Psi^{(2)}_n}{\partial\ct} = \hat{\ham}_n^{\mathrm{pert}}\Psi^{(2)}_n-\lambda\Psi^{(2)}_n\left(\rmi\frac{\partial_A\A^{(1)}\partial^A\Psi^{(1)}_n}{\Psi^{(1)}_n}+\frac{1}{2}\frac{\partial_A\partial^A\Psi^{(1)}_n}{\Psi^{(1)}_n}\right).
\end{equation}
The terms proportional to $\lambda$ are identified as the first quantum gravitational corrections. We follow the strategy introduced in \cite{Kiefer1991} and project these terms along the direction normal to the hypersurfaces of constant $S^{(0)}$. By using \eref{background-expansion-1}, \eref{background-expansion-2} and \eref{schroedinger-equation}, the quantum gravitational correction terms can be represented in the form  
\begin{eqnarray}
\label{eq:corrected-schroedinger}
\mathcal{V}_n^{\mathrm{QG}}&:=
- \frac{\lambda}{4}\,\mathrm{Re}\left[\frac{1}{\Psi^{(1)}_n\mathcal{P}}\left(\hat{\ham}_n^{\mathrm{pert}}\right)^2\Psi^{(1)}_n
+\rmi \frac{1}{\Psi^{(1)}_n}\left(\frac{\partial}{\partial\ct}\frac{\hat{\ham}_n^{\mathrm{pert}}}{\mathcal{P}}\right)\Psi^{(1)}_n\right].
\end{eqnarray}
We follow the treatment of \cite{Kiefer2012a,Brizuela2016,Brizuela2016a} and only take the real part of the corrections \eref{eq:corrected-schroedinger} in order to preserve unitarity defined with respect to the Schr\"odinger inner product on the Hilbert space of the perturbations. The question of unitarity in the context of the canonical approach to quantum gravity and the semiclassical expansion is controversially discussed and an interesting topic on its own, see e.g. \cite{Kiefer1991,Barvinsky1993a,Bertoni1996,Barvinsky2014,Kamenshchik2018a,Steinwachs2018, Kiefer2018, Chataignier2019}.
The term $\mathcal{V}_n^{\mathrm{QG}}$ in \eref{eq:corrected-schroedinger} might be viewed as a contribution to the effective potential
\begin{equation}
\rmi\frac{\partial\Psi^{(2)}_n}{\partial\ct} = -\frac{1}{2}\frac{\partial^2\Psi^{(2)}_{n}}{\partial q_n^2} +\mathcal{V}^{\mathrm{eff}}_{n}\Psi^{(2)}_{n},\qquad \mathcal{V}^{\mathrm{eff}}_{n} := \frac{1}{2}\omega_n^2q_n^2 + \mathcal{V}^{\mathrm{QG}}_{n}.\label{effpot}
\end{equation}
The iterative scheme of the semiclassical expansion implies that equations obtained at lower orders in $\lambda$ are used to derive equations arising at higher order in the expansion. 
In order to solve the corrected Schr\"odinger equation \eref{effpot} for $\Psi^{(2)}_n$, in addition knowledge about the solution $\Psi^{(1)}_n$ of the uncorrected Schr\"odinger equation \eref{schroedinger-equation}, which enters \eref{eq:corrected-schroedinger}, is required.
      
\section{Cosmological power spectra in the Schr\"odinger picture}
\label{sec:qg-corrections}

In order to extract physical information from the semiclassical expansion we need to relate observations to the \wkb{} states $\Psi_n^{(j)}$, where the $(j)$ indicates the order of the semiclassical expansion.
The inflationary perturbations are assumed to be Gaussian, which means that they are determined by the two-point correlation function. 
The main observable in inflationary cosmology is the inflationary power spectrum, which results from a Fourier transform of the two-point correlation function.
Since observational data do not show any evidence for non-Gaussian features we do not consider higher $n$-point correlation functions. 
This is consistent with our truncation of \eref{S2} to quadratic order in the perturbations, since investigations of e.g. the bispectrum would imply interaction terms cubic in the perturbations. 
In the Schr\"odinger picture, this suggests that the $\Psi_n^{(j)}$ obtained from the semiclassical expansion can be assumed to be normalized Gaussian states, 
\begin{equation}
\label{eq:wavefunction-ansatz}
\Psi_n^{(j)}(\ct;q_n)= N_n^{(j)}(\ct) \exp\left(-\frac{1}{2}\Omega_n^{(j)}(\ct) q_n^2\right),\quad \mathrm{Re}(\Omega_n)>0.
\end{equation}
They are fully characterized by the complex Gaussian width $\Omega_n(\ct)$ which depends parametrically on the semiclassical time $\ct$.
For Gaussian states \eref{eq:wavefunction-ansatz}, the quantum average in the \wkb{} state $\Psi_{\mathrm{pert}}^{(j)}=\prod_n\Psi_{n}^{(j)}$ can be evaluated explicitly. 
The two-point correlation function in the Schr\"odinger picture is a simple function of the Gaussian width \cite{Martin2012}, 
\begin{eqnarray}
\fl
\left\langle\Psi_{\mathrm{pert}}^{(j)} |v_\vc{k}v^\ast_\vc{k^{\prime}}| \Psi_{\mathrm{pert}}^{(j)} \right\rangle
= \frac{2\pi^2}{k^3}P^{(j)}_v\delta(\vc{k-k^{\prime}}),\qquad& P^{(j)}_v(\ct;k):= \frac{k^3}{4\pi^2}\text{Re}\left[\Omega_v^{(j)}(k;\ct)\right]^{-1},\label{PowerSpectrum}\\
\fl      
\left\langle\Psi_{\mathrm{pert}}^{(j)} |u_\vc{k}^Iu^{J\, \ast}_\vc{k^{\prime}}| \Psi_{\mathrm{pert}}^{(j)} \right\rangle
= \frac{2\pi^2}{k^3}P^{(j)}_{u}\delta(\vc{k-k^{\prime}})\delta^{IJ},
\qquad& P^{(j)}_u(\ct;k):= \frac{k^3}{4\pi^2}\text{Re}\left[\Omega_{u}^{(j)}(k;\ct)\right]^{-1}.\label{PowerSpectrum2}
\end{eqnarray}
Thus, knowledge of the $\Omega_n^{(j)}$ fully determines the power spectrum $P^{(j)}_v$ up to order $\mathcal{O}(\lambda^j)$ of the semiclassical expansion.      
Since the canonical field variables for the scalar and tensor perturbations $v$ and $u^{I}$ are related to the original perturbations $\delta\varphi_{\mathrm{gi}}$ and $h^{I}$ via \eref{eq:mukhanov-sasaki}, the corresponding power spectra are related to \eref{PowerSpectrum}-\eref{PowerSpectrum2} by
\begin{equation}
\label{eq:primordial-power-spectra}
P_{\rm S}(k) := \frac{1}{a^2z_{\rm S}^2}P_v(k),\qquad  P_{\rm T}(k) := \frac{2}{a^2z_{\rm T}^2}P_u(k).
\end{equation}
The extra factor of $2$ in $ P_{\rm T}(k)$ accounts for the two polarizations. For notational simplicity we suppressed the index $(j)$.
The power spectra can be parametrized by the power law forms
\begin{eqnarray}
P_{\rm S}(k) &= A_{\rm S}(k_\ast)\left(\frac{k}{k_\ast}\right)^{n_{\rm S}(k_\ast)-1+\ldots},\qquad P_{\rm T}(k) &= A_{\rm T}(k_\ast)\left(\frac{k}{k_\ast}\right)^{n_{\rm T}(k_\ast)+\ldots}.\label{Pscalar}
\end{eqnarray}
The pivot scale $k_*$ is chosen to correspond to a mode within the experimentally accessible window of scales that re-entered the horizon $N_{\mathrm{e}}$ efolds after the end of inflation. In terms of the parametrization \eref{Pscalar}, the power spectra are characterized by their amplitudes $A_{\mathrm{S}/\mathrm{T}}$ which measure the heights, and their spectral indices $n_{\mathrm{S}/\mathrm{T}}$, which measure the tilts 
\begin{eqnarray}
n_{\rm S}:=1+\frac{\mathrm{d}\log{P_S}}{\mathrm{d}\log{k}}\Big\vert_{k=k_{*}},\label{eq:scalar-index}\qquad &n_{\rm T}:=\frac{\mathrm{d}\log{P_T}}{\mathrm{d}\log{k}}\Big\vert_{k=k_{*}}.
\end{eqnarray}
Since the primordial tensor modes have not yet been measured there only exists an upper bound on $A_{\mathrm{T}}$ and it is convenient to introduce the tensor-to-scalar ratio
\begin{equation}
r=\frac{P_{\rm T}(k_{*})}{P_{\rm S}(k_{*})}.\label{tts}
\end{equation}
For single field models of inflation, there is a consistency equation relating $r$ and $n_{\mathrm{T}}$
\begin{equation}
r=-8n_{\mathrm{T}}\label{conseq}.
\end{equation}

\subsection{Power spectra without quantum gravitational corrections}
At order $\mathcal{O}(\lambda^0)$ of the semiclassical expansion, we obtain the Schr\"odinger equation \eref{schroedinger-equation} for the states $\Psi^{(1)}_n(\ct,q_n)$. According to \eref{eq:wavefunction-ansatz}, we insert the Gaussian ansatz 
\begin{equation}
\Psi_n^{(1)}= N_n^{(1)} \exp\left(-\frac{1}{2}\Omega_n^{(1)} q_n^2\right)\,\label{Gasusspsi1}
\end{equation}
into \eref{schroedinger-equation} and collect terms of equal order in the $q_n$. This leads to the two separate equations
\begin{eqnarray}
\rmi\frac{\mathrm{d}N^{(1)}  }{\mathrm{d} \ct}       = \frac{1}{2}\Omega^{(1)} N^{(1)},    \label{eq:app-uncorrected1}\qquad
\rmi\frac{\mathrm{d}\Omega^{(1)}}{\mathrm{d} \ct} = \left(\Omega^{(1)}\right)^2 - \omega^2. \label{eq:app-uncorrected2}
\end{eqnarray}
Here and in what follows we suppress the subindex $n$ which labels the different modes and types of perturbations. The frequency $\omega=\omega_{\mathrm{S}/\mathrm{T}}$ is given by \eref{perturbation-frequencies} for the scalar and tensor modes respectively.
The equation for $N_k^{(1)}$ just reproduces the usual normalization condition for the Gaussian. 
The equation for $\Omega^{(1)}_k$ is a first order non-linear differential equation.\footnote{Introducing the auxiliary variable $f^{(1)}$ via $\Omega^{(1)}:=-\rmi\partial_{\ct}\log (f^{(1)})$, \eref{eq:app-uncorrected2} acquires the structure of the equation for a harmonic oscillator with time-dependent frequency in accordance with the standard mode equation obtained in the Heisenberg quantization \cite{Martin2012}.} In accordance with \cite{Noh2001}, to linear order in the slow-roll approximation we find for the frequencies of the scalar and tensor modes
\begin{equation}
  \label{perturbation-frequencies-slow-roll}
  \omega^2_\mathrm{S}(\ct,k) = \omega^2_{\mathrm{DS}} - 3\frac{\Epsilon_\mathrm{S}}{\ct^2},
  \qquad
  \omega^2_\mathrm{T}(\ct,k) = \omega^2_{\mathrm{DS}}(\ct,k) - 3\frac{\Epsilon_\mathrm{T}}{\ct^2},
\end{equation}
where we have defined the abbreviations $\Epsilon_\mathrm{S}$ and $\Epsilon_\mathrm{T}$ which collect the contributions from the slow-roll parameters 
\begin{eqnarray}
\Epsilon_\mathrm{S}:=2\varepsilon_{1}-\varepsilon_{2}-\varepsilon_{3}+\varepsilon_{4},\qquad\Epsilon_\mathrm{T}:= \varepsilon_{1} + \varepsilon_{3},\label{eps}
\end{eqnarray}
and the universal time dependent De Sitter frequency 
\begin{equation}
\omega^2_{\mathrm{DS}}(k,\ct):=k^2-\frac{2}{\ct^2}\,.\label{DeSitterFreq}
\end{equation}	
The equation \eref{eq:app-uncorrected2} for $\Omega^{(1)}$ can be solved analytically. The result can be expressed in terms of Bessel functions, which in turn can be expanded in powers of the slow-roll parameters. However, since we work to first order in the slow-roll approximation, we might as well incorporate this directly into the solution by making the ansatz 
\begin{equation}
\Omega^{(1)}:= k(\Omega_{\mathrm{DS}} + \Epsilon \Omega_\Epsilon),\label{AnsatzOm1}
\end{equation}
with $\Epsilon=\Epsilon_{\mathrm{S}/\mathrm{T}}$, for scalar and tensor modes respectively. Inserting \eref{AnsatzOm1} together with \eref{perturbation-frequencies-slow-roll} and \eref{eps} into \eref{eq:app-uncorrected2}, changing the independent variable from $\ct$ to $x:=-k\ct$ and collecting terms of equal power in $\Epsilon$, we obtain the two differential equations
\begin{eqnarray}
\frac{\mathrm{d}\Omega_{\mathrm{DS}}}{\mathrm{d}x}&= \rmi\Omega_{\mathrm{DS}}^2 - \rmi\frac{x^2 - 2}{x^2},      \label{eq:uncorrected-se1}\\
\frac{\mathrm{d}\Omega_{\Epsilon}}{\mathrm{d}x}  &= 2\rmi\Omega_{\mathrm{DS}}\Omega_\Epsilon + \frac{3\rmi}{x^2}.  \label{eq:uncorrected-se2}
\end{eqnarray}
The system \eref{eq:uncorrected-se1} and \eref{eq:uncorrected-se2} can be solved successively by first solving for $\Omega_{\mathrm{DS}}$ and subsequently for $\Omega_{\Epsilon}$. In order to solve these differential equations, we choose the physical Bunch-Davies boundary condition, that is, we require that the solution $\Omega^{(1)}$ of \eref{eq:app-uncorrected2} matches the solution $\Omega_{\infty}^{(1)}$ of the equation obtained form \eref{eq:app-uncorrected2} in the limit $\ct\to-\infty$ (corresponding to $x\to\infty$). Since the frequencies $\omega_{\mathrm{S}/\mathrm{T}}$ become time-independent for $\ct\to-\infty$, the asymptotic limit of \eref{eq:app-uncorrected2} reads,
\begin{eqnarray}
\rmi\frac{\mathrm{d}\Omega^{(1)}_{\infty}}{\mathrm{d} \ct} = \left(\Omega^{(1)}_{\infty}\right)^2 - k^2. \label{eq:app-flat}
\end{eqnarray}
An obvious solution to \eref{eq:app-flat} is the time-independent Gaussian width $\Omega_{\infty}^{(1)}=k$. Therefore, in the limit $\ct\to-\infty$, the \wkb{} wave function $\Psi^{(1)}$, satisfies a stationary Schr\"odinger equation $\hat{\ham}^{\mathrm{pert}}\Psi^{(1)}=0$ with the Hamiltonian $\hat{\ham}^{\mathrm{pert}}$ of a harmonic oscillator with time-independent frequency $\omega_{\infty}=k$.
In view of the ansatz \eref{AnsatzOm1}, imposing the early-time asymptotic Bunch-Davies boundary condition
\begin{equation}
\lim_{\ct\to-\infty}\Omega^{(1)}(\ct)\equiv\Omega_{\infty}^{(1)}=k,\label{BDOm1}
\end{equation}
implies the asymptotic boundary conditions 
\begin{equation}
\lim_{x\to\infty}\Omega_{\mathrm{DS}}(x)=1,\qquad\lim_{x\to\infty}\Omega_{\mathcal{E}}(x)=0.\label{BDs}
\end{equation} 
With the boundary conditions \eref{BDs}, the solutions to \eref{eq:uncorrected-se1} and \eref{eq:uncorrected-se2} read
\begin{eqnarray}
\Omega_{\mathrm{DS}} &=  \frac{x^2 - \rmi x^{-1}}{ x^2 + 1 },\label{solDs}\\
\Omega_{\Epsilon}   &= \rmi\frac{1 + (2\rmi+x)x - 2e^{2\rmi x}{x^3}[\pi-\rmi \mathrm{Ei}(-2\rmi x)]}{x ( x - \rmi )^2}.\label{solEps}
\end{eqnarray}
Here $\mathrm{Ei}(z)$ is the exponential integral function of complex argument $z$ (defined on the complex plane with a branch cut along the negative $z$-axis), which is most conveniently defined in terms of the exponential integral function $E_{1}(z)$, which in turn is defined explicitly by its integral representation for $\mathrm{Re}(z)>0$,
\begin{eqnarray}
\mathrm{Ei}(z):=-\mathrm{E}_{1}(-z)+\frac{1}{2}\left[\ln\left(z\right)-\ln\left(\frac{1}{z}\right)\right]-\ln\left(-z\right),\\
 \mathrm{E}_{1}(z):=\int_{1}^{\infty}\mathrm{d}t\,\frac{e^{-zt}}{t},\qquad \mathrm{Re}(z)>0.
\end{eqnarray}
For small (large) arguments $z\ll1$ ($z\gg1$), the exponential integral $\mathrm{Ei}(z)$ has the (asymptotic) expansions
\begin{eqnarray}
\label{asexEi}
\fl \mathrm{Ei}(z)=\gamma_{\mathrm{E}}+\frac{1}{2}\left[\ln\left(z\right)-\ln\left(\frac{1}{z}\right)\right]+\sum_{k=1}^{\infty}\frac{1}{k!}\frac{z^k}{k},\quad |z|\ll1,\\
\fl
\mathrm{Ei}(z)=\frac{1}{2}\left[\ln\left(z\right)-\ln\left(\frac{1}{z}\right)\right]-\ln\left(-z\right)+\frac{1}{z}e^{-z}\left[1+\mathcal{O}\left(\frac{1}{z}\right)\right],\quad|z|\gg1.
\end{eqnarray}  
The logarithms in the conversion between $\mathrm{Ei}$ and $\mathrm{E}_{1}$ arise because $\mathrm{Ei}$ is multivalued. 
Combining \eref{solDs} and \eref{solEps} with the ansatz \eref{AnsatzOm1} yields the solution for the Gaussian width $\Omega^{(1)}$ to first order in the slow-roll approximation.
According to \eref{PowerSpectrum} and \eref{eq:primordial-power-spectra}, knowledge of $\mathrm{Re}\left(\Omega^{(1)}\right)$ fully determines the inflationary scalar and tensor power spectra. Making use of the expansion \eref{asexEi}, at superhorizon scales $k^{-1}\gg\chubble^{-1}$ (corresponding to $x\ll1$), we obtain
\begin{equation}
\text{Re}(\Omega^{(1)}) \approx kx^2 \left[1 - 2c_{\gamma}\Epsilon+2\Epsilon\log x)\right].\label{SHLReOm1}
\end{equation}
Here, $c_\gamma:=2 - \gamma_{\rm E} -\log2\approx0.7296$ is a numerical constant involving the Euler-Mascheroni constant $\gamma_{\rm E}\approx0.5772$.
From \eref{PowerSpectrum} and \eref{eq:primordial-power-spectra}, we obtain the scalar and tensor power spectra
\begin{eqnarray}
P^{(1)}_{\mathrm{S}}(k) &\approx \frac{1}{(2\pi z_{\mathrm{ S}})^2}\left(\frac{k}{ax}\right)^2\left[1 + 2c_{\gamma}\Epsilon_{\mathrm{S}}-2\Epsilon_{\mathrm{S}}\log x)\right],\label{FinPS1S}\\
P^{(1)}_{\mathrm{T}}(k) &\approx \frac{2}{(2\pi z_{\mathrm{ T}})^2}\left(\frac{k}{ax}\right)^2\left[1 + 2c_{\gamma}\Epsilon_{\mathrm{T}}-2\Epsilon_{\mathrm{T}}\log x)\right].\label{FinPS1T}
\end{eqnarray}
Using the explicit expressions \eref{eps} for $\Epsilon_{\mathrm{S}/\mathrm{T}}$ and the first order slow-roll relation between the conformal time and the conformal Hubble parameter
\begin{equation}
\label{eq:conformal-time-slow-roll}
\ct\approx-\frac{1+\varepsilon_1}{\chubble},
\end{equation} 
as well as the explicit expression \eref{eq:z} for $z_{\mathrm{S}/\mathrm{T}}^2$ expressed in terms of slow-roll parameters,
\begin{eqnarray}
z_{\mathrm{S}}^2=4U\frac{\varepsilon_{1}+\varepsilon_{3}}{(1+\varepsilon_{3})^2},\qquad z_{\mathrm{T}}^2=\frac{U}{2},
\end{eqnarray}
we obtain the inflationary power spectra to first order in the slow-roll approximation 
\begin{eqnarray}
\label{eq:uncorrected-scalar-spectrum}
P^{(1)}_{\mathrm{S}}(k) \approx& \frac{W}{96 \pi^2(\varepsilon_{1}+\varepsilon_{3})}
\Big[1 - \frac{1}{3}\left(5\varepsilon_{1} -6c_{\gamma}\Epsilon_{\mathrm{S}}\right)-2\Epsilon_{\mathrm{S}}\log\left(k/\chubble\right)\Big],\\
P^{(1)}_{\mathrm{T}}(k)\approx& \frac{W}{6\pi^2}
\left[1-\frac{1}{3}\left(5\varepsilon_1+6\varepsilon_3-6c_{\gamma}\Epsilon_{\mathrm{T}}\right)-2\Epsilon_{\mathrm{T}}\ln(k/H)\right]\label{TPS1}.
\end{eqnarray}
Note that instead of \eref{eq:attractor1}, we have used the relation including terms linear in the slow-roll parameters
\begin{equation}
\frac{\chubble^2}{Ua^2}\approx \frac{W}{6}\left(1+\frac{\varepsilon_1}{3}-2\varepsilon_3\right),\label{WSL}
\end{equation}
in order to express factors of $\chubble^2/Ua^2$ in \eref{eq:uncorrected-scalar-spectrum} and \eref{TPS1} in terms of $W$.
Equation \eref{WSL} follows from combining equations \eref{eq:friedmann-equation} and \eref{eq:eom-hubble} in the slow-roll approximation.
Finally, using \eref{eq:scalar-index} and \eref{tts}, we obtain the spectral observables from \eref{eq:uncorrected-scalar-spectrum} and \eref{TPS1}, 
\begin{eqnarray}
\label{eq:uncorrected-spectral-index}
n_{\mathrm{S}}^{(1)} =& 1 - 2\Epsilon_{\mathrm{S}} = 1 - 2\left(2\varepsilon_{1} - \varepsilon_{2} - \varepsilon_{3} +\varepsilon_{4}\right),\\
n_{\mathrm{T}}^{(1)} =& -2\Epsilon_{\mathrm{T}} = -2\left(\varepsilon_{1} + \varepsilon_{3}\right),\\
r^{(1)}=&\frac{A_{\mathrm{T}}^{(1)}}{A_{\mathrm{S}}^{(1)}}= 16\left(\varepsilon_{1} + \varepsilon_{3}\right)=-8n_{\mathrm{T}}^{(1)}.\label{runcorrected}
\end{eqnarray}
These expressions for the inflationary observables coincide with the expressions found in \cite{Noh2001} within the standard Heisenberg quantization of the fluctuations propagating in the classical time-dependent \flrw{} background.\footnote{Note that $\varepsilon_{1}$ and $\varepsilon_2$ in \cite{Noh2001} are defined with signs opposite to our definitions \eref{HubbleSLP} and \eref{PotSLP}.}
In contrast, here the observables \eref{eq:uncorrected-spectral-index}-\eref{runcorrected} were derived in the Schr\"odinger picture by performing the semiclassical expansion of the Wheeler-DeWitt equation. This result provides an important consistency check for the method, as it shows that the semiclassical expansion of the Wheeler-DeWitt equation oder-by-order recovers the classical background equations \eref{eq:friedmann-equation}-\eref{eq:reduced-scalar-eom} as well as the Schr\"odinger equation for the fluctuations propagating on this classical background \eref{schroedinger-equation}, which leads to the correct inflationary slow-roll observables \eref{eq:uncorrected-spectral-index}. Finally, the next order in the semiclassical expansion yields the first quantum gravitational corrections \eref{effpot}, whose impact on the inflationary observables we derive in the next section.

\subsection{Power spectra including quantum gravitational corrections}
In this subsection, we calculate the contribution of the first quantum gravitational corrections to the inflationary power spectra of the general scalar-tensor theory \eref{sec-infldyn:action}, which is obtained at order $\mathcal{O}\left(\lambda^1\right)$ of the semiclassical expansion of the Wheeler-DeWitt equation \eref{wdw}. The \wkb{} state $\Psi^{(2)}_n$ at order $\mathcal{O}\left(\lambda^1\right)$ is defined by the corrected Schr\"odinger equation \eref{effpot}. The quantum gravitational correction term \eref{eq:corrected-schroedinger} requires knowledge of the \wkb{} states $\Psi^{(1)}_n$ obtained at order $\mathcal{O}\left(\lambda^0\right)$ as the solution of the uncorrected Schr\"odinger equation \eref{schroedinger-equation}. Under the assumption that the $\Psi^{(1)}_n$ have the Gaussian form \eref{Gasusspsi1}, the quantum gravitational corrections \eref{eq:corrected-schroedinger} can be expressed as a function of the background potential $\mathcal{P}$, as well as the frequency $\omega^2_n$ and the Gaussian width $\Omega^{(1)}_n$ of the scalar and tensor modes, respectively. Suppressing again the subindex labeling different species and modes, the quantum gravitational corrections  \eref{eq:corrected-schroedinger} read  
\begin{eqnarray}
\mathcal{V}^\mathrm{QG}=\mathrm{Re}&
  \left\{-\frac{\Omega^{(1)}\left[\Omega^{(1)}-2\rmi\partial_{\ct}(\ln\mathcal{P})\right]
         +2\left[\left(\Omega^{(1)}\right)^2-\omega^2\right]}{16\mathcal{P}}\right.\nonumber\\
&-\left.\frac{\left[\rmi\partial_{\ct}(\ln\mathcal{P})-3\Omega^{(1)}\right]\left[\left(\Omega^{(1)}\right)^2-\omega^2\right]
         +2\rmi\omega\partial_{\ct}\omega}{8\mathcal{P}}\right.q^2\nonumber\\
&-\left.\frac{\left[(\Omega^{(1)})^2-\omega^2\right]^2}{16\mathcal{P}}q^4\right\}.\label{QGCorrectOm1}
\end{eqnarray}
Inserting \eref{QGCorrectOm1} into the corrected Schr\"odinger equation \eref{effpot}  with the Gaussian ansatz
\begin{equation}
\Psi^{(2)}=N^{(2)}\mathrm{exp}\left(-\frac{1}{2}\Omega^{(2)}q^2\right)
\end{equation}  
and collecting terms of equal power in $q$ yields an equation for $\Omega^{(2)}$\footnote{The terms independent of $q$ only enter the equation for the normalization factor. In addition, since we assumed that at each order the \wkb{} wave function is of the Gaussian form \eref{eq:wavefunction-ansatz}, we neglect the terms proportional to $q^4$ in \eref{QGCorrectOm1}. This is consistent with our approach, as we have quantized the Hamiltonian \eref{eq:hamiltonian-constraint}, including only up to quadratic terms in the perturbations, not including interactions among the perturbations. This truncation might also be justified on a phenomenological basis, as so far there is no observational evidence for primordial non-Gaussianities, see e.g. \cite{Ade2016c,Renaux-Petel2015}. Therefore, the assumption that the perturbations are in their vacuum (Gaussian) state seems to be a reasonable one. 
},
\begin{equation}
  \label{eq:uncorrected-reduced-schroedinger}
 \rmi\frac{\mathrm{d}\Omega^{(2)}}{\mathrm{d}\ct} = \left(\Omega^{(2)}\right)^2 - \left(\omega^2-\omega^2_{\mathrm{QG}}\right).
\end{equation}
The frequency $\omega_{\mathrm{QG}}$ which includes the quantum gravitational corrections is defined as
\begin{eqnarray}
  \label{eq:corrected-frequency}
  \omega_{\mathrm{QG}}^2:=\left.\frac{\partial^2\mathcal{V}^{\mathrm{QG}}}{\partial q^2}\right|_{q=0}=& \frac{\mathrm{Re}\left(\Omega^{(1)}\right)}{4\mathcal{P}}\Big\{\mathrm{Im}\left(\Omega^{(1)}\right)\left[3\mathrm{Im}\left(\Omega^{(1)}\right)-2\partial_{\ct}\left(\ln \mathcal{P}\right)\right]\nonumber\\
  &-3\left[\mathrm{Re}\left(\Omega^{(1)}\right)^2-\mathrm{Im}\left(\Omega^{(1)}\right)^2-\omega^2\right]\Big\}. 
\end{eqnarray}
The inhomogeneous non-linear ordinary first order differential equation \eref{eq:uncorrected-reduced-schroedinger} is difficult to solve analytically. Since we expect the quantum gravitational contributions to $\Omega^{(2)}$ to be small, we linearize \eref{eq:uncorrected-reduced-schroedinger} around $\Omega^{(1)}$,
\begin{equation}
\delta\Omega:=\Omega^{(2)}-\Omega^{(1)},\qquad \delta\Omega/\Omega^{(1)}\ll1.\label{splitOm2}
\end{equation}
Upon using \eref{eq:app-uncorrected2} linearization of \eref{eq:uncorrected-reduced-schroedinger} leads to
\begin{equation}
\label{eq:perturbations}
\rmi\frac{\mathrm{d}\delta\Omega}{\mathrm{d}\ct} = 2\Omega^{(1)}\delta\Omega  +\omega_{\mathrm{QG}}^2.
\end{equation}
In the following we assume in addition that the quantum gravitational contributions are small compared to the slow-roll contributions $\propto\Epsilon\Omega_{\Epsilon}$ to the uncorrected Gaussian width $\Omega^{(1)}$, that is, we assume $\delta\Omega/\Omega^{(1)}\ll|\varepsilon_i|\ll1$. In particular, this implies that we only keep the dominant De Sitter contributions, corresponding to terms $\mathcal{O}\left(\varepsilon_{i}^{0}\right)$, in the quantum gravitational frequency $\omega_{\mathrm{QG}}^2$ and the De Sitter part $\Omega_{\mathrm{DS}}$ in the solution $\Omega^{(1)}$ obtained in the previous order of the expansion \eref{solDs}. For completeness, we have worked out the observational consequences following from including the first order slow-roll contributions $\mathcal{O}\left(\varepsilon_i^1\right)$ to the quantum gravitational corrections in \ref{Appendix}.  Focusing on the dominant De Sitter contributions to $\omega^2_{\mathrm{QG}}$, we obtain
\begin{eqnarray}
\label{eq:corrected-frequencyDeSitter}
\omega_{\mathrm{QG}}^2 \approx \frac{\mathrm{Re}\left(\Omega_{\mathrm{DS}}\right)}{4\mathcal{P}}&\Bigg\{\mathrm{Im}\left(\Omega_{\mathrm{DS}}\right)\left[3\mathrm{Im}\left(\Omega_{\mathrm{DS}}\right)-2\partial_{\ct}\left(\ln \mathcal{P}\right)\right]\nonumber\\
&-3\left[\mathrm{Re}\left(\Omega_{\mathrm{DS}}\right)^2-\mathrm{Im}\left(\Omega_{\mathrm{DS}}\right)^2-\omega^2_{\mathrm{DS}}\right]\Bigg\}.
\end{eqnarray}
Next, we express derivatives with respect to conformal time $\ct$ by $\partial_{\ct}=-k\partial_x$, insert \eref{DeSitterFreq}, \eref{AnsatzOm1}, \eref{solDs} and make use of the background equations of motion in the slow-roll approximation \eref{eq:attractor2} to express the potential $\mathcal{P}$ in terms of the constant conformal Hubble parameter in De Sitter space  $H$ and the non-minimal coupling $U$, neglecting any time dependence to zeroth order in slow-roll. In this way, we finally arrive at
\begin{equation}
\omega_{\mathrm{QGDS}}^2 \approx-\frac{W}{144 k}\frac{x^4(x^2-11)}{(x^2+1)^3}.\label{dscontqgcorr}
\end{equation} 
In order to solve \eref{eq:perturbations} with the source \eref{dscontqgcorr}, we again impose the asymptotic Bunch-Davis boundary condition for $\Omega^{(2)}$, that is,  we require that the Gaussian width $\Omega^{(2)}$ is time-independent in the limit $\ct\to-\infty$ (i.e. $x\to\infty$), for which \eref{eq:uncorrected-reduced-schroedinger} reduces to the algebraic condition
\begin{eqnarray}
\left(\Omega^{(2)}_{\infty}\right)^2=k^2+\frac{W}{144 k}.
\end{eqnarray}
Here we have made use of the expansion $x\gg1$ in \eref{dscontqgcorr}, to find the asymptotic limit
\begin{eqnarray}
\lim_{x\to\infty}\omega^2_{\mathrm{QGDS}}=-\frac{W}{144 k}.
\end{eqnarray}
With \eref{BDOm1}, this implies for $\delta\Omega_{\infty}$,
\begin{eqnarray}
\delta\Omega_{\infty}=\frac{\Omega^{(1)}_{\infty}}{2}\left[\left(\frac{\Omega^{(2)}_{\infty}}{\Omega^{(1)}_{\infty}}\right)^2-1\right]=\frac{W}{288 k^2 }.\label{bcpert}
\end{eqnarray}
The linearized ordinary first order equation for the De Sitter part of the quantum gravitational corrections in terms of the independent variable $x=-k\ct$ then reads
\begin{eqnarray}
\frac{\mathrm{d}}{\mathrm{d} x}\delta\Omega_{\mathrm{DS}}= 2\rmi\left(\frac{x^2 - \rmi x^{-1}}{ x^2 + 1 }\right)\delta\Omega_{\mathrm{DS}}-\rmi\frac{W}{144 k^2}\frac{x^4(x^2-11)}{(x^2+1)^3}.\label{deSitterlin}
\end{eqnarray}
With the boundary condition \eref{bcpert}, the solution reads
\begin{eqnarray}
\label{soldeltaO}
\delta\Omega_{\mathrm{DS}}=\frac{W}{288k^2}\frac{\rme^{2 \rmi x}x^2}{(x-\rmi)^2}&\Bigg[3 \rmi\pi\frac{3+\rme^4}{\rme^2}+\rme^{-2\rmi x}\frac{1+x\left(x-6\rmi\right)}{(x+\rmi)^2}\nonumber\\
&+3\rme^2\mathrm{Ei}(-2-2\rmi x)+9\rme^{-2}\mathrm{Ei}(2-2\rmi x)\Bigg].
\end{eqnarray}
The power spectrum \eref{PowerSpectrum} obtained from the Gaussian width $\Omega^{(2)}$ is approximated as
\begin{equation}
P^{(2)}= \frac{k^3}{2\pi^2}\frac{1}{2\text{Re}\left[\Omega^{(1)}+\delta\Omega_{\mathrm{DS}}\right]}\approx P^{(1)}\left[1-\frac{\mathrm{Re}\left(\delta\Omega_{\mathrm{DS}}\right)}{\mathrm{Re}\left(\Omega^{(1)}\right)}\right].\label{PS}
\end{equation}
In order to calculate the impact of the De Sitter contributions to the quantum gravitational corrections on the inflationary power spectra, all we need is to extract the superhorizon limit $(x\ll1)$ of the real part of the solution \eref{soldeltaO},
\begin{equation}
\mathrm{Re}\left(\delta\Omega_{\mathrm{DS}}\right)=\frac{\beta_0 W}{144}\frac{x^2}{k^2 }+\mathcal{O}\left(x^4\right)\approx-\frac{W}{ 72}\frac{x^2}{k^2 },\label{QGCPS}
\end{equation}
with the numerical constant $\beta_0:=[1-3\rme^2\mathrm{Ei}(-2)-9\rme^{-2}\mathrm{Ei}(2)]/2\approx -2$.
The De Sitter contribution to $\Omega^{(1)}$ in the superhorizon limit reads
\begin{equation}
\mathrm{Re}\left(\Omega^{(1)}\right)\approx\mathrm{Re}\left(\Omega_{\mathrm{DS}}\right)\approx kx^2.\label{SHOmDS}
\end{equation}
Evaluating the ratio $\mathrm{Re}\left(\delta\Omega_{\mathrm{DS}}\right)/\mathrm{Re}\left(\Omega^{(1)}\right)$ by combining \eref{QGCPS} with \eref{SHOmDS}, we obtain
\begin{equation}
\fl
P^{(2)}(k)=P^{(1)}(k)\Big[1+\delta_{\mathrm{QGDS}}(k/k_0)\Big],\qquad \delta_{\mathrm{QGDS}}(k/k_0):=\frac{W}{ 72}\left(\frac{k_0}{k }\right)^3.\label{QGC}
\end{equation}
In the last step, we have re-introduced the reference wavelength $k_0=l_0^{-1}$ originating from reversing the rescalings \eref{rescal1} and \eref{rescal2}. Note in particular that the uncorrected part of the power spectrum $P^{(1)}$ is invariant under this rescaling due to its logarithmic dependence on the invariant ratio $k/\chubble$ -- only the quantum  gravitational corrections are affected.
Using the definitions of the scalar and tensor power spectra \eref{eq:primordial-power-spectra} together with the results obtained in the previous order of the expansion \eref{FinPS1S} and \eref{FinPS1T}, we finally obtain the corrected scalar and tensor power spectra as
\begin{eqnarray}
\fl
P^{(2)}_{\mathrm{S}}(k)\approx \frac{W}{96 \pi^2(\varepsilon_{1}+\varepsilon_{3})}
\Big[1 - \frac{1}{3}\left(5\varepsilon_{1}-6c_{\gamma}\Epsilon_{\mathrm{S}}\right)-2\Epsilon_{\mathrm{S}}\log\left(k/\chubble\right)+\delta_{\mathrm{QGDS}}(k/k_0)\Big],\label{P2SQG}\\
\fl
P^{(2)}_{\mathrm{T}}(k)\approx\frac{W}{6\pi^2}\left[1-\frac{1}{3}\left(5\varepsilon_1+6\varepsilon_3-6c_{\gamma}\Epsilon_{\mathrm{T}}\right)-2\Epsilon_{\mathrm{T}}\ln(k/H)+\delta_{\mathrm{QGDS}}(k/k_0)\right].\label{P2TQG}
\end{eqnarray}
In deriving \eref{P2SQG} and \eref{P2TQG}, we have again neglected any slow-roll contributions to the quantum gravitational corrections and only kept the dominant De Sitter contribution \eref{QGC}. We discuss the particular features and observable consequences of the corrected power spectra \eref{P2SQG} and \eref{P2TQG} in more detail in the next section.

\section{Observational signatures of quantum gravitational effects}
\label{sec:observational-signatures}
In this section, we discuss several features of the quantum gravitational corrections to the power spectra, their observable signatures, their magnitude and the impact of the non-minimal coupling.

\begin{enumerate}
\item Just as the uncorrected power spectra \eref{eq:uncorrected-scalar-spectrum}, \eref{TPS1}, the quantum gravitational corrected power spectra \eref{P2SQG} and \eref{P2TQG} become time-independent at superhorizon scales. This ``freezing'' is important as it allows to calculate the power spectrum in the superhorizon limit at horizon crossing. 

\item The dominant De Sitter part of the quantum gravitational corrections $\delta_{\mathrm{QGDS}}$ is universal, in the sense that it equally contributes to the scalar and tensor power spectrum.
In particular, this implies that these corrections will drop out in the tensor-to-scalar ratio as found for the minimally coupled case \cite{Brizuela2016}. This degeneracy between the scalar and tensorial spectra can be broken by including slow-roll contributions to the quantum gravitational corrections. Although these slow-roll contributions are additionally suppressed by powers of the slow-roll parameters and therefore even less relevant than the already minuscule dominant De Sitter part of the quantum gravitational corrections, the analysis is of theoretical interest and for completeness included in \ref{Appendix}. 

\item The quantum gravitational effects lead to an enhancement of the power spectra. 

\item The quantum gravitational corrections have a characteristic $1/k^3$-dependence, which allows -- at least in principle -- to observationally distinguish between the quantum gravitational contributions and the uncorrected constant De Sitter and logarithmic $k$-dependent slow-roll parts of the power spectrum. 

\item The quantum gravitational corrections are heavily suppressed relative to the uncorrected part of the power spectra. Due to the $(k_0/k_*)^3$ dependence, it is clear that the quantum gravitational corrections are strongest on the largest scales (smallest values of $k_*$) and also depend on the infrared regularizing reference scale $k_0\sim\ell_0^{-1}$. The latter is undetermined a priori. Note, that the uncorrected part of the power spectra is independent of the reference scale $k_0$.
 
More precise statements about the magnitude of the quantum gravitational corrections can be made by adopting the standard power law parametrization of the power spectra \eref{Pscalar}, which allows to characterize the quantum gravitational corrected power spectra \eref{P2SQG} and \eref{P2TQG} by their amplitudes and spectral indices,
\begin{eqnarray}
A_{\mathrm{S}}^{(2)}\approx \frac{W_*}{96 \pi^2(\varepsilon_{1}+\varepsilon_{3})}
\Big[1 - \frac{1}{3}\left(5\varepsilon_{1}-6c_{\gamma}\Epsilon_{\mathrm{S}}\right)+\delta_{\mathrm{QGDS}}(k_*/k_0)\Big],\label{As}\\
A_{\mathrm{T}}^{(2)}\approx \frac{W_*}{6\pi^2}\left[1-\frac{1}{3}\left(5\varepsilon_1+6\varepsilon_3-6c_{\gamma}\Epsilon_{\mathrm{T}}\right)+\delta_{\mathrm{QGDS}}(k_*/k_0)\right],\\
n_{\mathrm{S}}^{(2)}\approx
1- 2\left(2\varepsilon_{1} - \varepsilon_{2} - \varepsilon_{3} +\varepsilon_{4}\right)-3\delta_{\mathrm{QGDS}}(k_*/k_0),\\
n_{\mathrm{T}}^{(2)}\approx
-2\left(\varepsilon_{1} + \varepsilon_{3}\right)-3\delta_{\mathrm{QGDS}}(k_*/k_0),\\
r^{(2)}\approx\frac{A_{\mathrm{T}}^{(2)}}{A_{\mathrm{S}}^{(2)}}\approx r^{(1)}=16(\varepsilon_1+\varepsilon_3)\neq -8 n_{\mathrm{T}}^{(2)}.\label{rtensor}
\end{eqnarray}
Here, $W_*$ denotes $W$ evaluated at the moment ${k_*=\chubble_*}$, when the pivot mode $k_{*}$ first crosses the horizon. Note that according to \eref{rtensor}, the quantum gravitational corrections lead to a tiny violation of the consistency condition \eref{conseq}.
The power-law ansatz \eref{Pscalar} is usually justified by the weak logarithmic scale dependence of the power spectra. While this is true for the uncorrected part, in view of the $1/k^3$ dependence of the quantum gravitational corrections, it seems questionable whether such a parametrization is adequate.
The pivot scale is chosen within the window of scales observable in the \cmb{} \cite{Akrami2018}, 
\begin{equation}
\fl
k^{\mathrm{min}}_*<k_*<k^{\mathrm{max}}_*,\qquad k^{\mathrm{min}}_*=10^{-4}\; \mathrm{ Mpc}^{-1},\qquad k^{\mathrm{max}}_*=10^{-1}\; \mathrm{ Mpc}^{-1}.\label{Rangek}
\end{equation}
Measurements of the \cmb{} constrain $A_{\mathrm{S}}$ and $n_{\mathrm{S}}$ and give an upper bound on the tensor-to-scalar ratio $r$, here quoted for $k_*=0.05\;\mathrm{Mpc}^{-1}$ \cite{Akrami2018},
\begin{eqnarray}
 A_{\mathrm{S},*}^{\mathrm{obs}}=\left(2.099\pm0.014\right)\times10^{-9},\quad& 68 \%\;\mathrm{CL},\label{Planck18As}\\
n_{\mathrm{S},*}^{\mathrm{obs}}=0.9649\pm 0.0042,\quad& 68 \%\;\mathrm{CL},\label{Planck18ns}\\
r^{\mathrm{obs}}_{*}<0.11,\quad& 95 \%\;\mathrm{CL}.\label{Planck18r}
\end{eqnarray}
The upper bound on $r^{\mathrm{obs}}_{*}$ is connected to an upper bound on the energy scale during inflation. This corresponds to an upper bound on the energy density, given by the potential $\hat{V}=M_{\mathrm{P}}^4W/4$, the potential in the Einstein frame, see. e.g. \cite{Steinwachs2018} for a discussion of different field parametrizations in scalar-tensor theories. Expressing $\hat{V}_{*}$ in terms of the inflationary observables yields
\begin{eqnarray}
\hat{V}_*=\left(\frac{M_{\mathrm{P}}^4W_*}{4}\right)\approx \frac{3}{2}\pi^2M_{\mathrm{P}}^4A_{\mathrm{T},*}^{(2)}\approx \frac{3}{2}\pi^2M_{\mathrm{P}}^4A_{\mathrm{S},*}^{(2)}\,r^{(2)}_{*}.\label{EscalInf}
\end{eqnarray}
Identifying $A^{(2)}\approx A^{(1)}\approx A^{\mathrm{obs}}$ and $r^{(2)}\approx r^{(1)}\approx r^{\mathrm{obs}}$, \eref{Planck18As} and \eref{Planck18r} imply the upper bound\footnote{It would be also interesting to compare the energy scale of inflation, including quantum gravitational corrections, with the energy scale derived from quantum cosmology \cite{Barvinsky1994a,Barvinsky2010,Calcagni2014}.}
\begin{eqnarray}
\frac{W_*}{72}\approx\frac{\left(6\pi^2 A_{\mathrm{S},*}^{\mathrm{obs}}\,r^{\mathrm{obs}}_{*}\right)}{72}\lesssim  10^{-10}.\label{estimate}
\end{eqnarray}	
The estimate \eref{estimate} shows that independently of the concrete scalar-tensor theory, i.e. independent of the choices for $U$, $G$, and $V$ in \eref{sec-infldyn:action}, for a viable inflationary model the quantum gravitational corrections to the power spectrum are suppressed relative to the uncorrected part by 
\begin{equation}
\delta_{\mathrm{QGDS}}(k_*/k_0)\lesssim  10^{-10}\left(\frac{k_0}{k_*}\right)^3.
\end{equation}
Therefore, the only way to enhance the impact of the quantum gravitational corrections is to increase the ratio $k_0/k_*$. 
While the quoted measurements \eref{Planck18As}-\eref{Planck18r} are obtained at a fixed pivot point $k_*=0.05\,\mathrm{Mpc}^{-1}$, this reference scale is in principle arbitrary and only constrained to lie with the interval \eref{Rangek} which is accessible to observations of the \cmb{}.
Since the leading quantum gravitational correction $\delta_{\mathrm{QGDS}}$ is strongest for the smallest value of $k_{*}$, experiments should be most sensitive to a detection of a potential quantum gravitational effect at $k_{*}^{\mathrm{min}}$.
Although the infrared scale $k_0$ is unspecified, an upper bound on $k_0$ might be derived along the lines of the discussion in \cite{Brizuela2016a}.
On the one hand, the observed scalar amplitude of the perturbations has a measured value $A_{\mathrm{S},*}^{\mathrm{obs}}$ with experimental uncertainty $\delta_{\mathrm{exp}}\approx0.014$.
On the other hand, the dominant quantum  gravitational correction to the amplitude reads $A_{\mathrm{S}}^{(2)}=A_{\mathrm{S}}^{(1)}\left(1+\delta_{\mathrm{QGDS}}\right)$.
In order for a quantum gravitational effect to be detectable we must have $\delta_{\mathrm{QGDS}}>\delta_{\mathrm{exp}}$.
Conversely, not having detected a quantum gravitational effect implies an upper bound on $k_{0}$ from $\delta_{\mathrm{QGDS}}\leq\delta_{\mathrm{exp}}$.
Using \eref{QGC} and \eref{EscalInf} to expressing $\delta_{\mathrm{QGDS}}$ in terms of observable quantities, we obtain the inequality
\begin{equation}
k_0\leq\left[\frac{72}{6\pi^2 A_{\mathrm{S},*}^{\mathrm{obs}}}\frac{\delta_{\mathrm{exp}}}{r_{*}^{\mathrm{obs}}}\right]^{1/3}k_{*}.\label{k0kstar}
\end{equation}
An upper bound $k_0^{\mathrm{max}}$ for $k_0$ is obtained for equality in \eref{k0kstar}. It largely depends on the ratio $\delta_{\mathrm{exp}}/r_{*}^{\mathrm{obs}}$. 
On the one hand, increasing experimental precision will lower $\delta_{\mathrm{exp}}$ and therefore lower $k_{0}^{\mathrm{max}}$. 
On the other hand, since the tensor-to-scalar ratio $r_{*}^{\mathrm{obs}}$ is only bounded from above, a measurement of $r_{*}^{\mathrm{obs}}$ smaller than the upper bound \eref{Planck18r} would increase $k_0^{\mathrm{max}}$.
In order to get a rough estimate of the order of magnitude, we insert values from \eref{Planck18As}-\eref{Planck18r}.
With $\delta_{\mathrm{exp}}=0.014$ and $r_{*}^{\mathrm{obs}}=0.11$ at $k_{*}=0.05$, we obtain
\begin{equation}
k_{0}^{\mathrm{max}}\approx 21 \,\mathrm{Mpc}^{-1}.\label{k0ksvalues}
\end{equation}  
If we had used the higher bound $\delta_{\mathrm{exp}}=1$, we would have obtained $k_{0}^{\mathrm{max}}\approx87\,\mathrm{Mpc}^{-1}$ for a pivot point $k_0=0.05\,\mathrm{Mpc}^{-1}$, in agreement with the estimate obtained in \cite{Brizuela2016a}.
The length scale corresponding to \eref{k0ksvalues} is $l_0^{\mathrm{max}}=1/k_{0}^{\mathrm{max}}\approx0.048\,\mathrm{Mpc}$.
For reference scales  $l_0<l_0^{\mathrm{max}}$ quantum gravitational effects would be resolvable within the assumed precision.
Let us compare this to the natural choice for the infrared cutoff, the radius of the observable universe, corresponding to a scale $k_0\approx k^{\mathrm{min}}_{*}$. 
Since the maximum value for the quantum gravitational effects is obtained for $k_*=k^{\mathrm{min}}_{*}$, the ratio is $\left(k_0/k_{*}\right)^3=1$.
Assuming that $A_{\mathrm{S},*}^{\mathrm{obs}}$ and $r^{\mathrm{obs}}_{*}$ do not change drastically under a change of the pivot point from $k_{*}=0.05\,\mathrm{Mpc}^{-1}$ to $k_{*}=k^{\mathrm{min}}_{*}=10^{-4}\,\mathrm{Mpc}^{-1}$, the dominant quantum gravitational corrections to the power spectra is of order 
\begin{equation}
\delta_{\mathrm{QGDS}}\approx 10^{-10}.
\end{equation}
\item
Finally, as a consistency check of our result, we should recover the results obtained in \cite{Brizuela2016} for the minimally coupled case. 
Indeed, inserting the constant non-minimal coupling function  $U(\varphi)=M_{\mathrm{P}}^2/2$ and $V=\mathcal{V}$ into $W=V/U^2$, the results $\eref{P2SQG}$ and $\eref{P2TQG}$ reduce to the corresponding expressions obtained in \cite{Brizuela2016}. 
Therefore, the main impact of the non-minimal coupling $U$ on the quantum gravitational corrections corresponds to a replacement of the constant Planck mass by the ``effective field dependent Planck mass'' ${M_{\mathrm{P}}^2\to M_{\mathrm{P}}^2(\varphi)=2U(\varphi)}$ -- a result that might have been expected naively. 

However, despite the arbitrariness in the field dependent non-minimal coupling $U$, for any viable scalar-tensor theory of inflation, the ratio $W=V/U^2$ is constrained by observations and therefore sets an upper bound on the magnitude of the dominant quantum corrections. 
Thus, for any scalar-tensor theory of the form \eref{sec-infldyn:action} which is consistent with observational data, the non-minimal coupling $U$ does not lead to an enhancement of the quantum gravitational corrections.
Nevertheless, the impact of the generalized potentials $U$, $G$ and $V$ enters in the subleading slow-roll contribution to the quantum gravitational corrections, discussed in \ref{Appendix}.
\end{enumerate}

\section{Conclusion}
\label{sec:conclusion}

In this paper we have calculated the first quantum gravitational corrections to the inflationary power spectra for a general scalar-tensor theory from a semi-classical expansion of the Wheeler-DeWitt equation.

Let us summarize the different steps and assumptions used in our analysis: 
We expanded the general scalar-tensor action around a flat \flrw{} background up to quadratic order in the perturbations.
After a Fourier transformation of the inhomogeneous perturbations, we proceeded with the canonical quantization of the combined background and perturbation variables.
The quantization of the Hamilton constraint lead to the Wheeler-DeWitt equation which describes the exact quantum dynamics.
We then performed a semiclassical expansion of the Wheeler-DeWitt equation based on a combined Born-Oppenheimer and \wkb{}-type approximation.
The Born-Oppenheimer approximation relies on the division of the configuration space variables into heavy and light degrees of freedom. In the cosmological context the background variables are naturally identified with the heavy degrees of freedom, while the Fourier modes of the perturbations are identified with the light degrees of freedom.  
At lowest order in the semiclassical expansion, we recovered the classical homogeneous background equations of motion and the notion of a background-dependent semiclassical time from the timeless Wheeler-DeWitt equation. At the next order, we obtained the Schr\"odinger equation for the perturbations evolving with respect to this semiclassical time.

Finally, at the subsequent order in the expansion, we derived the first quantum gravitational contributions which can be represented in the form of a corrected Schr\"odinger equation.
In order to extract the observational consequences, we calculated the inflationary power spectra by assuming that at each order of the expansion the semiclassical wavefunction has a Gaussian form and satisfies the asymptotic Bunch-Davies boundary condition at early times. For such Gaussian states, the inflationary power spectra are fully determined by the real part of the Gaussian width. The assumption about the Gaussian nature of the wave function is natural if the system is in the ground state. For a recent extension to excited states, see \cite{Brizuela2019}.

Under these assumptions, at the level of the uncorrected Schr\"odinger equation, we recovered the standard power spectra, usually obtained in the Heisenberg quantization of the perturbations propagating on a classical background. In contrast, in our approach these results follow directly from the semiclassical expansion of the Wheeler-DeWitt equation. This shows that the semiclassical expansion not only correctly reproduces the classical background equations, but also the inflationary power spectra. Therefore, this provides an important consistency check for the semiclassical treatment to the geometrodynamical approach to quantum gravity.
Finally, from the corrected Schr\"odinger equation, we derived the first quantum gravitational corrections to the inflationary power spectra. Since these quantum gravitational corrections are highly suppressed, we restricted our analysis in the main text to the dominant De Sitter contribution. We derived the first order slow-roll contributions to the quantum gravitational corrections separately in the appendix.

We found that the dominant quantum gravitational corrections for a general scalar-tensor theory leads to an increase in the amplitude of the inflationary scalar and tensor power spectra. This increase is universal in the sense that it affects both power spectra in the same way. However, compared to the uncorrected part of the power spectra,  we found that even the dominant quantum gravitational corrections are strongly suppressed. This is in agreement with previous results obtained for a minimally coupled scalar field \cite{Kiefer2012a,Brizuela2016,Brizuela2016a}. Although the non-minimal coupling $U$ enters the quantum gravitational corrections to the power spectra, it only enters in the dimensionless combination $W=V/U^2$. Since the uncorrected power spectra are already proportional to $W$, observations put strong constraints on the value $W_*\lesssim10^{-9}$ at horizon crossing and therefore on the magnitude of the quantum gravitational corrections. This implies that independently of the concrete choice for the generalized potentials $U$, $G$ and $V$ present in the general scalar-tensor theory \eref{sec-infldyn:action}, the dominant quantum gravitational corrections are strongly suppressed as long as the observational constraints for a successful phase of inflation are satisfied.
In particular, this shows that it is not possible to enhance the quantum gravitational corrections to the inflationary power spectra due to the presence of a non-minimal coupling. 
The impact of the generalized potentials on the quantum gravitational corrections only affects the subleading slow-roll contributions. 

Independently of the strong suppression factor, the quantum gravitational corrections feature a characteristic scale dependence $\propto 1/k^3$ which has been found in similar approaches \cite{Kiefer2012a,Kamenshchik2013,Brizuela2016,Brizuela2016a,Kamenshchik2016a, Kamenshchik2018}. This scale dependence is not only a prediction of quantum gravity but also provides an observational signature. In fact, the scale dependence of the quantum gravitational corrections enters the power spectra in the form $(k_0/k_*)^3$ where $k_*$ is the pivot point and $k_0$ the infrared regulating scale which arises in the flat \flrw{} universe.
Although the quantum gravitational corrections are suppressed by $W_*/72\approx10^{-10}$, depending on the values for $k_0$ and $k_*$, the scaling factor $(k_0/k_*)^3$ might increase the magnitude of the quantum gravitational corrections. While, the value of $k_{*}$ is constrained to lie within the observable window, the value of $k_0$ is a priori undetermined. Quantum gravitational effects are strongest for a pivot point at the lower end of the allowed interval $k_{*}=k_{*}^{\mathrm{min}}\approx 10^{-4}\,\mathrm{Mpc}^{-1}$. A natural choice for the infrared regulating scale $l_0$ is the size of the observable universe, which corresponds to a scale $k_0=k_{*}^{\mathrm{min}}$. For these choices of $k_*$ and $k_0$, the ratio $(k_0/k_*)^3$ is of order one. In this case, quantum gravitational effects are suppressed by $10^{-10}$ and therefore unobservable. Conversely, for quantum gravitational effects to come into observational reach a value $k_0\approx10^{-1}\,\mathrm{Mpc}^{-1}$ would be required. This, in turn, would single out a preferred astrophysical length scale $l_0\approx 10 \,\mathrm{Mpc}$. However, since the reference scale $l_0$ was introduced to regularize the infinite spatial volume integral arising in a homogeneous and isotropic flat \flrw{} universe, such a value seems to be inconsistent as it is well below the smoothing scale of $l_{\mathrm{smooth}}\gtrsim 200\,\mathrm{Mpc}$. We therefore conclude that within the available precision of the current observations quantum gravitational from the semiclassical expansion of the Wheeler-DeWitt equation are unobservable -- even for general scalar-tensor theories.

\section*{Acknowledgements}
M.~W.~was supported by grant GRK 2044 of the German Research Foundation (DFG).

\newpage
\appendix

\section{Slow-roll contributions to the quantum gravitational corrections}
\label{Appendix}
In this appendix we derive the subleading quantum gravitational corrections to first order in the slow-roll parameters.

There are two reasons why we include this analysis: first, due to the universal character of the dominant De Sitter contribution to the quantum gravitational corrections, which affects both scalar and tensor modes in the same way, it is interesting to see whether this degeneracy is lifted upon inclusion of the first slow-roll contributions.
In particular, it allows to investigate the impact of the quantum gravitational corrections on the tensor-to-scalar ratio, which remains unaffected by the dominant universal De Sitter contribution.
Second, the slow-roll contributions to the quantum gravitational corrections also carry information about the generalized potentials $U$, $G$ and $V$, which characterize the general scalar-tensor theory \eref{sec-infldyn:action} and therefore allow to test the dependence of the quantum gravitational corrections $\delta_{\mathrm{QG}}(k/k_0)$ on the parameters of the theory such as e.g. a non-minimal coupling -- of course the uncorrected part of the power spectrum is sensitive to the generalized potentials via the dependence on the generalized slow-roll parameters. 

While the following analysis is based on the assumption that the quantum gravitational corrections are small $\delta\Omega/\Omega^{(1)}\ll1$ and that the slow-roll approximation is valid $|\varepsilon_i\ll1|$, we keep mixed terms of the form $~\delta\Omega\varepsilon$, while neglecting terms $\delta\Omega^2$ and $\varepsilon_i^2$ (for a complete treatment, the higher order slow-roll contributions to the uncorrected power spectra up to the order where they compete with the quantum gravitational slow-roll corrections should be included, which we neglect however). Under these assumptions, we have to solve the linearized equation \eref{eq:perturbations} with the terms linear in the slow-roll parameters included 
\begin{equation}
\label{eq:de-width-perturbation}
\frac{\mathrm{d}}{\mathrm{d} x}\delta\Omega = 2\rmi(\Omega_{\mathrm{DS}} + \Epsilon\Omega_{\Epsilon})\delta \Omega
- \rmi\frac{W}{144k^2}(\omega^2_{0}+\omega_{1}^2\varepsilon_1+\omega_3^2\varepsilon_3+ \Epsilon\omega^2_\Epsilon ).
\end{equation}
Compared to the equation for the dominant De Sitter contributions \eref{deSitterlin}, in \eref{eq:de-width-perturbation} we include the slow-roll contribution $\Epsilon\Omega_{\Epsilon}$ in the linear term as well as in the quantum gravitational frequency \eref{eq:corrected-frequencyDeSitter}, which we have parametrized in terms of the four functions 
\begin{eqnarray}
\omega_0^2(x)      &:=\frac{144k^2}{W}\omega^2_{\mathrm{QGDS}}= x^4\frac{x^2-11}{(1+x^2)^3},\\
\omega_1^2(x)
&:=-\frac{2}{3}x^4\frac{11x^2-49}{(1+x^2)^3},\\
\omega_{3}^2(x)
&:=-8x^4\frac{x^2-5}{(1+x^2)^3},\\
\omega_{\Epsilon}^2(x)	   &:=-\frac{x^4}{(1+x^2)^4}\left[P(x)+Q(x)+\overline{Q}(x)\right],
\end{eqnarray}
with the polynomials (the bar denotes complex conjugation)
\begin{eqnarray}
\fl
P(x) = 7x^6-21x^4+89x^2-27,\\
\fl
Q(x) = e^{2\rmi x}(\rmi+x)^4\left[6x^4-34x^2+11-\rmi(20x^3-22x)\right]\left[\rmi\pi+\mathrm{Ei}(-2\rmi x)\right].
\end{eqnarray}

\begin{figure}
	\centering
		\includegraphics[width=0.75\textwidth]{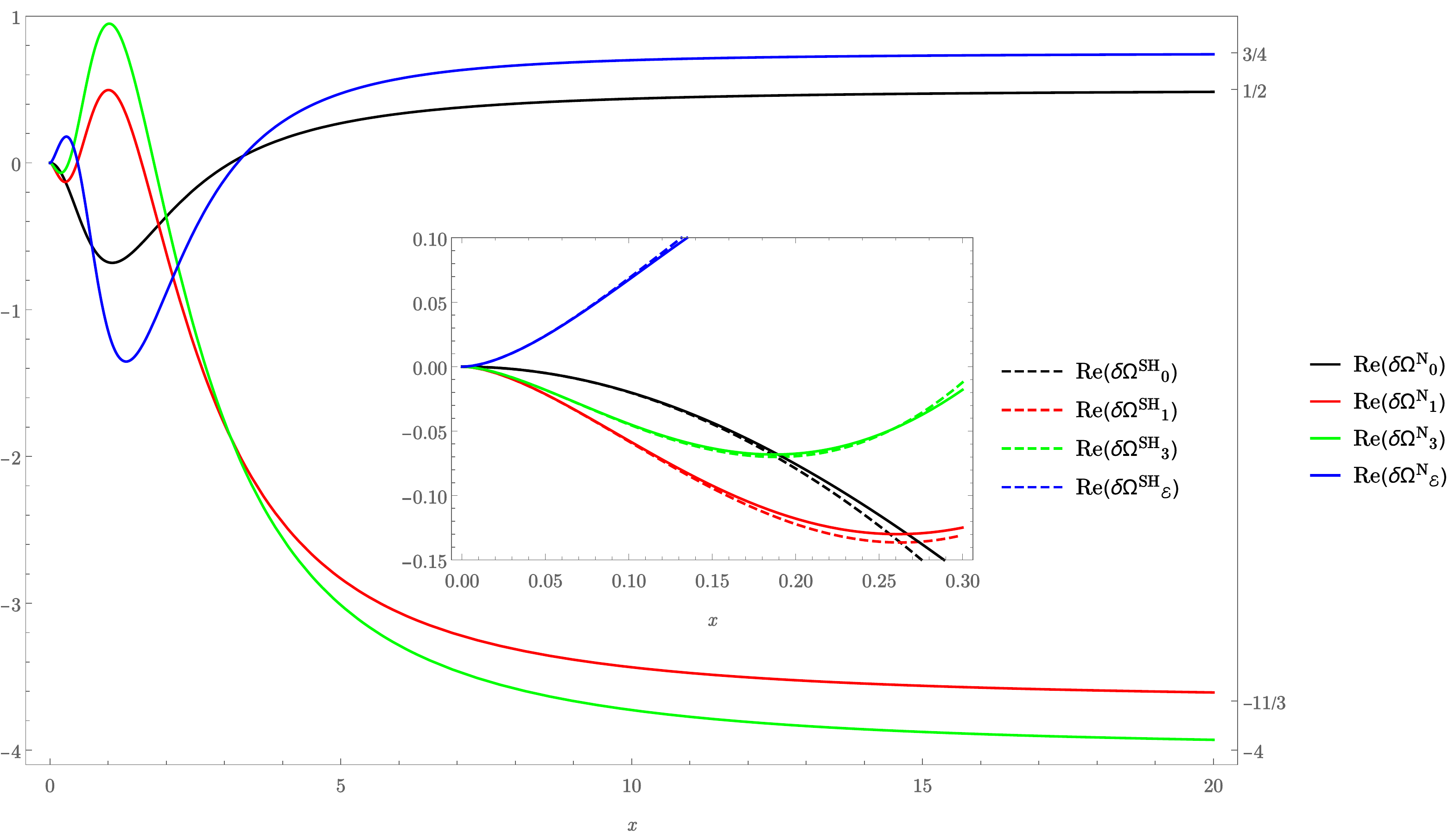}
		\caption{Numerical solutions to \eref{eq:linearised-corrections-de}. The inlay compares the superhorizon behavior of the functions with their analytical approximations \eref{SHsol}.}
		\label{fig:numerical_plot}
\end{figure}

We make the following ansatz for the perturbation of the Gaussian width:
\begin{equation}
\label{eq:delta-omega}
\delta\Omega = \frac{W}{144k^2}
\left(\delta\Omega_0+ \varepsilon_{1}\delta\Omega_1+\varepsilon_{3}\delta\Omega_3 + \Epsilon\delta\Omega_\Epsilon \right).
\end{equation}
Equation \eref{eq:de-width-perturbation} can then be written as a system of linear equations:
\begin{equation}
\label{eq:linearised-corrections-de}
\frac{\mathrm{d}\delta\Omega_i}{\mathrm{d} x}(x)
=
M_{i}{}^{j}(x)\delta\Omega_j(x) - X_i(x),
\end{equation}
with the matrix $M$ and the vector $X$ defined as
\begin{equation}
M(x):=
2\rmi\left(
\begin{array}{cccc}
\Omega_{\mathrm{DS}}    &    0     & 0        & 0 \\
\rmi/x  & \Omega_{\mathrm{DS}}  & 0        & 0 \\
\rmi/x         &    0     & \Omega_{\mathrm{DS}}  & 0 \\
\Omega_\Epsilon       &    0     & 0        & \Omega_{\mathrm{DS}} \\
\end{array}
\right)
,\qquad
X:=
\rmi\left(
\begin{array}{c}
\omega^2_0\\
\omega^2_1\\
\omega^2_3\\
\omega^2_\Epsilon
\end{array}\right).
\label{Matrix}
\end{equation}
We again impose the asymptotic Bunch-Davis boundary condition.
Up to linear order in the slow-roll parameters, in the limit $\ct\to-\infty$, \eref{eq:uncorrected-reduced-schroedinger} reduces to
\begin{equation}
(\Omega_\infty^{(2)})^2=k^2+\frac{W}{144k}\left[1-\frac{22}{3}\varepsilon_1-8\varepsilon_3+\frac{3}{2}\Epsilon\right].
\end{equation}
Using \eref{BDOm1} and \eref{eq:delta-omega} implies the asymptotic values
\begin{equation}
\delta\Omega^\infty_i=\left(\frac{1}{2},-\frac{11}{3},-4,\frac{3}{4}\right).\label{asymptDO}
\end{equation}
The solution to \eref{eq:linearised-corrections-de} that satisfies this asymptotic condition can formally be written as
\begin{eqnarray}
\label{eq:qgc-solution}
\delta\Omega_i(x)
=
\mathfrak{M}_{i}{}^{j}(x)
\left\{\delta\Omega^{\infty}_{j}-\int^{x}_{\infty} \left[\mathfrak{M}^{-1}\right]_{j}{}^{k}(z)X_k(z)\,\mathrm{d} z\right\},\\
\mathfrak{M}(x):=\left[\exp\left(\int_{\infty}^x M(y)\,\mathrm{d} y\right)\right]_{ij},
\end{eqnarray}
Due to the non-diagonal elements of the matrix exponential the solutions $\delta\Omega_1$, $\delta\Omega_3$, and $\delta\Omega_\Epsilon$ cannot be computed analytically, and one has to resort to numerical methods.
For the numerical evaluation of \eref{eq:linearised-corrections-de} a finite cutoff $x_0$ for the lower integration bound has to be chosen. For sufficiently large values of $x_0$, the final results for the numerical solutions $\delta\Omega_i^{\mathrm{N}}$ do not depend on this choice as they quickly asymptote their constant values \eref{asymptDO}, but in order to facilitate the comparison of our results with \cite{Brizuela2016a}, we fixed the numerical value to $x_0=10^6$.
The real part of the $\delta\Omega_i^{\mathrm{N}}$, required for the calculation of the power spectra, are plotted in Fig. \ref{fig:numerical_plot}.
The inflationary power spectra are obtained in the superhorizon limit $x\ll1$. In principle, this limit can be obtained directly from the numerical solutions $\delta\Omega^{\mathrm{N}}_i$. However, we can use a hybrid analytic-numerical approach to extract the analytic $x$-dependence of the power spectra in the superhorizon limit.
In the super-horizon regime $x\ll1$,
the system of linear equations \eref{eq:linearised-corrections-de} reduces to the simple set of equations
\begin{eqnarray}
\fl
\frac{\mathrm{d}\delta\Omega_0^{\mathrm{SH}}}{\mathrm{d}x}=2x^{-1}\delta\Omega_0^{\mathrm{SH}},\qquad
&\frac{\mathrm{d}\delta\Omega_{\Epsilon}^{\mathrm{SH}}}{\mathrm{d}x}=2x^{-1}\left(\delta\Omega_{\Epsilon}^{\mathrm{SH}}+\delta\Omega_0^{\mathrm{SH}}\right),\\
\fl
\frac{\mathrm{d}\delta\Omega_1^{\mathrm{SH}}}{\mathrm{d}x}=2x^{-1}\left(\delta\Omega_1^{\mathrm{SH}}-\delta\Omega_0^{\mathrm{SH}}\right),\qquad
&\frac{\mathrm{d}\delta\Omega_3^{\mathrm{SH}}}{\mathrm{d}x}=2x^{-1}\left(\delta\Omega_3^{\mathrm{SH}}-\delta\Omega_0^{\mathrm{SH}}\right),
\end{eqnarray}
with the real part of the solutions determined up to integration constants $\beta_i$,
\begin{eqnarray}
\fl
\mathrm{Re}\left(\delta\Omega_0^{\mathrm{SH}}\right) = \beta_0\,x^2,\qquad &\mathrm{Re}\left(\delta\Omega_{\Epsilon}^{\mathrm{SH}}\right)= (\beta_\Epsilon+2\beta_0\log x)x^2,\nonumber\\
\fl
\mathrm{Re}\left(\delta\Omega_{1}^{\mathrm{SH}}\right)=(\beta_1-2\beta_0\log x)x^2,
\qquad&
\mathrm{Re}\left(\delta\Omega_{3}^{\mathrm{SH}}\right)=(\beta_3-2\beta_0\log x)x^2.\label{SHsol}
\end{eqnarray}
Since the superhorizon solutions $\delta\Omega^{\mathrm{SH}}_{i}$ in \eref{SHsol} are obtained from \eref{eq:linearised-corrections-de} in the limit $x\ll1$, the integration constants $\beta_0$ and $\beta_i$, $i=1,3,\Epsilon$ cannot be determined by the asymptotic Bunch-Davies boundary conditions \eref{asymptDO}, which are imposed at $x\gg1$. 
Instead, they have to be determined by fitting the analytic superhorizon solutions $\delta\Omega^{\mathrm{SH}}_{i}$ in \eref{SHsol} to the numerical solutions $\delta\Omega^{\mathrm{N}}_{i}$ at $x\ll1$:
\begin{eqnarray}
\beta_0\approx -1.98,\qquad
\beta_1\approx 3.30,\qquad 
\beta_3    \approx 4.62,\qquad
\beta_\Epsilon\approx -2.24.\label{fit}
\end{eqnarray}
Here $\beta_0$ consistently reproduces the constant in the analysis of the dominant De Sitter contribution in \eref{QGCPS}.
Using the analytical solutions $\delta\Omega^{\mathrm{SH}}_{i}$ in \eref{SHsol} with the fits \eref{fit} for the coefficients $\beta_i$, the power spectra including the slow-roll contributions to the quantum gravitational corrections are compactly written as 
\begin{eqnarray}
P^{(2)}_{\rm S/T} \approx P^{(1)}_{\rm S/T} \left( 1 + \delta^{\rm S/T}_{\mathrm{QG}}\right).\label{PSCorrected}
\end{eqnarray}
The quantum gravitational corrections $\delta^{\rm S/T}_{\mathrm{QG}}$ for the scalar and tensorial perturbations which include the slow-roll contributions are
\begin{equation}
\fl
\delta^{\mathrm{S}/\mathrm{T}}_{\mathrm{QG}}\approx\delta_{\mathrm{QGDS}}\left[1+\beta_0^{-1}\beta_1\varepsilon_1+\beta_0^{-1}\beta_3\varepsilon_3+\left(\beta_0^{-1}\beta_\Epsilon+2c_\gamma\right)\Epsilon_{\mathrm{S}/\mathrm{T}}-2(\varepsilon_1+\varepsilon_3)\log{x}\right).\label{QGSL}
\end{equation}
where we recall that $\delta_{\mathrm{QGDS}}$ is given by
\begin{equation}
\delta_{\mathrm{QGDS}}(k/k_0) = -\frac{\beta_0 W}{144}\left(\frac{k_0}{k}\right)^3 \approx \frac{W}{72}\left(\frac{k_0}{k}\right)^3.
\end{equation}
In contrast to the dominant universal De Sitter part of the quantum gravitational corrections, as can be inferred form \eref{QGSL}, the subleading slow-roll contributions to the quantum gravitational corrections are different for scalar and tensor perturbations.
The spectral observables can be obtained straightforwardly by inserting the corrected power spectra \eref{PSCorrected} into \eref{eq:scalar-index}.  
In particular, this breaks the degeneracy for the tensor-to-scalar ratio, defined in \eref{tts}, which was insensitive to the quantum gravitational corrections in the pure De Sitter case.

We also compare our results to the analysis performed in \cite{Brizuela2016a} for a single minimally coupled scalar field with a canonically normalized kinetic term and an arbitrary scalar potential $\mathcal{V}$.
While the results of \cite{Brizuela2016} for the dominant De Sitter contributions to the quantum gravitational corrections $\delta_{\mathrm{QGDS}}$ agree with our results for the identification $\mathcal{V}=M_{\mathrm{P}}^4W/4$ with $U=M_{\mathrm{P}}^2/2$ (for which $\mathcal{V}=V$), we find differences in the subleading slow-roll contributions to the quantum gravitational corrections considered in \cite{Brizuela2016a}. First, when comparing our results with the results of \cite{Brizuela2016a}, care is required as we have parametrized our final result for the corrected power spectra in terms of $W$, while the results in \cite{Brizuela2016a} were expressed in terms of $H^2/M_{\mathrm{P}}^2$. The conversion from $H^2/M_{\mathrm{P}}^2$ to $W$ induces additional terms linear in the slow-roll parameters, see \eref{WSL}.  Second, there is a true difference between both results which originates from the treatment of the ansatz \eref{eq:delta-omega} in the differential equation \eref{eq:de-width-perturbation}. We took into account derivatives of $W$, responsible for the off-diagonal terms $i/x$ in the matrix \eref{Matrix}, while the authors of \cite{Brizuela2016a} expanded $W$ around horizon crossing $W_*$ in \eref{eq:de-width-perturbation} and attributed the additional terms linear in slow-roll to the source term in \eref{eq:de-width-perturbation}. Consequently, they assumed a constant value $W_*$ in the ansatz \eref{eq:delta-omega}. Both procedures ultimately lead to different results. In particular, the implementation of boundary data becomes more complicated in the procedure followed \cite{Brizuela2016a}, as the corrections to the Gaussian width $\delta\Omega_{i}$ do not asymptote to constant values at $x\to\infty$  without further modifications -- in contrast to our approach, where this arises naturally \eref{asymptDO}.

\section*{References}

\bibliography{references}{}
\bibliographystyle{ieeetr}

\end{document}